\documentclass[preprint,NumberedRefs]{JASA}
\pdfoutput=1

\usepackage{textcomp}

\usepackage{bm}
\usepackage{graphicx}

\DeclareMathAlphabet{\mathsfbr}{OT1}{cmss}{m}{n}
\SetMathAlphabet{\mathsfbr}{bold}{OT1}{cmss}{bx}{n}
\DeclareRobustCommand{\msf}[1]{%
  \ifcat\noexpand#1\relax\msfgreek{#1}\else\mathsfbr{#1}\fi
}

\makeatletter
\newcommand{\msfgreek}[1]{\csname s\expandafter\@gobble\string#1\endcsname}
\makeatother

\DeclareRobustCommand{\mcal}[1]{%
  \ifcat\noexpand#1\relax\mathnormal{#1}\else\cal{#1}\fi
}
\DeclareRobustCommand{\BM}[1]{%
  \ifcat\noexpand#1\relax\bm{\boldUppercaseItalicGreek{#1}}\else\bm{#1}\fi
}
\makeatletter
\newcommand{\boldUppercaseItalicGreek}[1]{\csname var\expandafter\@gobble\string#1\endcsname}
\makeatother
\newcommand{\rv}[1]{\msf{#1}}
\newcommand{\RV}[1]{\bm{\msf{#1}}}

\newcommand{\V}[1]{\bm{#1}}
\newcommand{\M}[1]{\BM{#1}}

\providecommand{\ist}{\hspace*{.3mm}}
\providecommand{\rmv}{\hspace*{-.3mm}}
\providecommand{\rrmv}{\hspace*{-1mm}}
\providecommand{\nn}{\nonumber}

\DeclareMathAlphabet{\mathpzc}{OT1}{pzc}{m}{it}

\begin{document}

\title[]{Bayesian Detection and Tracking of Odontocetes in 3-D from Their Echolocation Clicks}


\author{Junsu Jang}
\author{Florian Meyer}
\author{Eric R. Snyder}
\author{Sean M. Wiggins}
\author{Simone Baumann-Pickering}
\author{John A. Hildebrand}

\affiliation{Scripps Institution of Oceanography, University of California San Diego, La Jolla, CA 92093, USA}

\email{\text{jujang@ucsd.edu}}




\begin{abstract}
Localizing and tracking of marine mammals can reveal key insights into behaviors underwater that otherwise would remain unexplored. A promising nonintrusive approach to obtaining location information of marine mammals is based on recordings of bio-acoustic signals by volumetric hydrophone arrays. Time-difference-of-arrival (TDOA) measurements of echolocation clicks\cite{Zim:B11} emitted by odontocetes can be extracted and used for detection, localization, and tracking in 3-D. We propose a data processing chain that automatically detects and tracks multiple odontocetes in 3-D from their echolocation clicks. First, TDOA measurements are extracted with a generalized cross-correlation that whitens received acoustic signals based on instrument noise statistics. Subsequently, odontocetes are tracked in the TDOA domain using a graph-based multi-target tracking (MTT) method to reject false TDOA measurements and close gaps of missed detections. The resulting TDOA estimates are then used by another graph-based MTT stage that estimates odontocete tracks in 3-D. The tracking capability of the proposed data processing chain is demonstrated on real acoustic data provided by two volumetric hydrophone arrays that recorded echolocation clicks from Cuvier's beaked whales (\textit{Ziphius cavirostris}). Simulation results show that the presented 3-D MTT method can outperform an existing approach that relies on hand annotation.
\end{abstract}

\maketitle



\section{Introduction}
\label{sec:introduction}

Passive acoustic monitoring (PAM) is a nonintrusive and efficient approach for studying and monitoring acoustically active animals, especially species that are challenging to observe visually. PAM enables detection, localization, and tracking and is therefore well-suited for studying abundance\cite{MarThoMarMelWarMorHarTya:J13, HilBauFraTriMerWigMcdGarHarMarTho:J15}, behaviors\cite{GasWigHil:J15}, and response to anthropogenic activities\cite{ThoBlaConKimMarThoOedHarBro:J20, KruRicFraReeTriSimRyaWigDenSchBau:J21}. With continuously increasing human activities in the ocean\cite{FalSchWatDerZerAndMorMor:J17, BuxMcKMenFriCroAngWit:J17, ErbMarSchRenSmiTriEmb:J19}, consistent monitoring and assessing the population, behavior, phenology, and physiology of marine organisms are necessary for making informed conservation plans and management policies\cite{PirBooCosFleKraLusMorNewSchWeiWelHar:J19}. Of particular monitoring interest are cetaceans since they are apex predators and environment sentinels\cite{HazAbrBroCarJacSavScaSydBog:J19}. Their animal density and geographic location can help understand complex environmental changes, e.g., caused by anthropogenic disturbances.

\begin{figure}[t]
   \includegraphics[width=\linewidth]{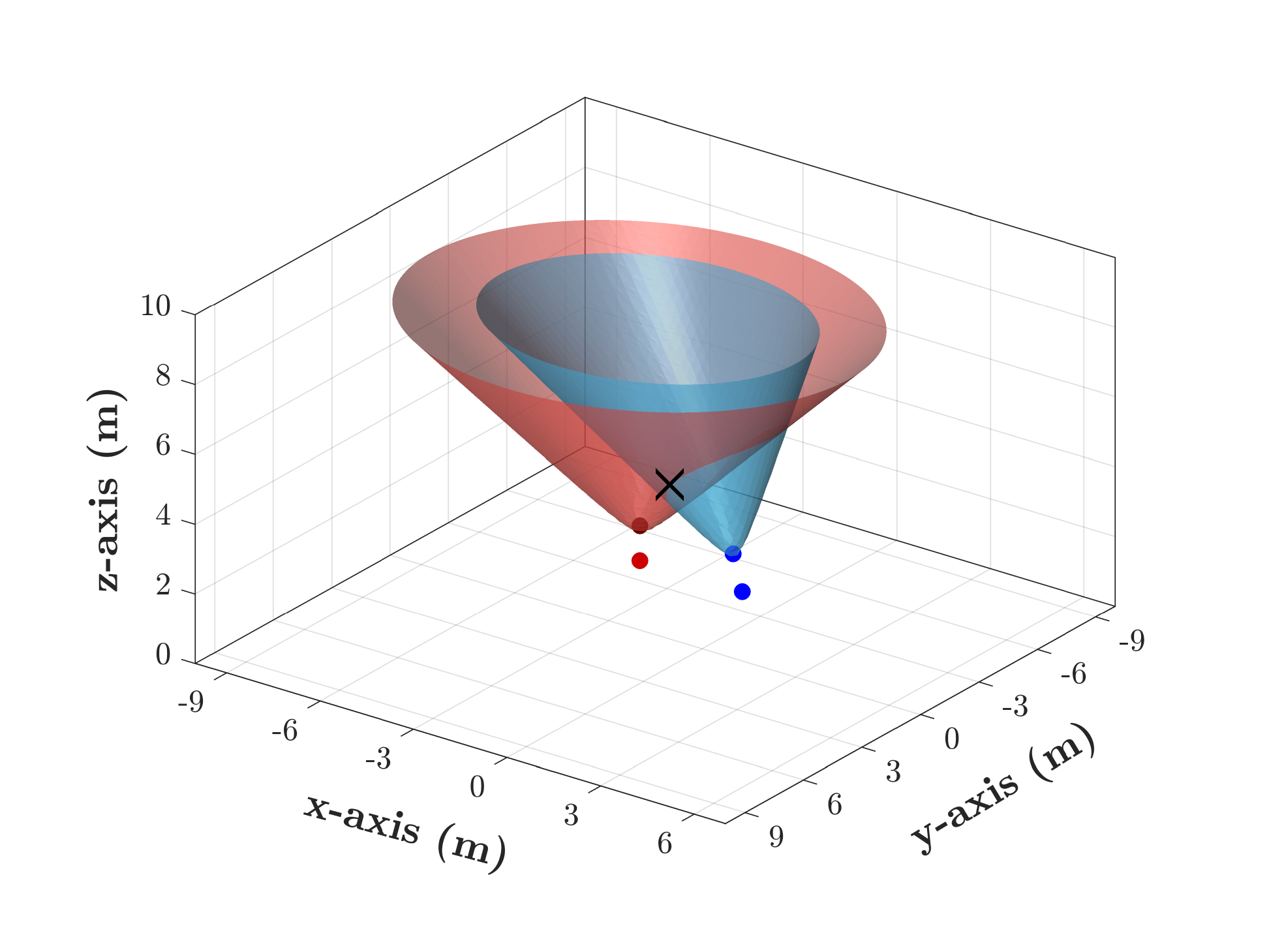}
   \caption{Example of acoustic source localization in 3-D from TDOA measurements. The cross is the location of the acoustic source. Each pair of red and blue dots represents a hydrophone pair that produces a TDOA measurement. Each TDOA measurement gives rise to a hyperboloid. The line of hyperboloid intersection specifies potential source locations. To obtain a unique source location, further hydrophone pairs are needed.}
   \label{fig:hyperboloids}
\end{figure}

Cetaceans are known to produce various types of sounds for communication, navigation, and foraging \cite{Zim:B11}. They are divided into two suborders: \textit{Odontocetes} (toothed whales) and \textit{Mysticetes} (baleen whales). Odontocetes predominantly use high-frequency whistles or burst pulses to communicate\cite{AuLam:B14}, while mysticetes produce low-frequency tonal calls, which when used in a pattern is considered song\cite{Clark:B90}. To locate prey and relevant features of the environment, odontocetes also emit echolocation clicks\cite{Zim:B11}, which are intense and directional short pulses. Since clicks are impulsive and broadband in frequency as well as frequently used underwater by odontocetes, they are promising signals for researchers to localize and track the echolocating odontocetes with PAM.

Various PAM technologies suitable for studying whales have been developed. Promising sensing approaches for PAM include towed arrays\cite{GruNosOle:J21}, fiber optic cables\cite{BouTawKriRorPotLanJohBreHauSchSto:J22}, mobile hydrophone recorders\cite{BarGriKliHar:J18,FreHarMatMelBarBauKli:J20}, and bottom-mounted hydrophone arrays\cite{GasWigHil:J15,HenMarManMat:J16}. Emerging accessible and inexpensive PAM technologies are expected to provide acoustic datasets that are orders of magnitude larger than datasets provided by conventional technologies \cite{GibBroGloJon:J19}. Thus, establishing effective algorithmic solutions for data processing, data management, data analysis, and performance evaluation is crucial. 

This work focuses on the acoustic data recorded by volumetric hydrophone arrays that can provide location information of echolocating odontocetes in a 3-D Euclidean space. Here, human operators are typically required to manually inspect acoustic data, make decisions on the presence of whales, and select promising measurements \cite{TenThoLamConKim:J22, BarNosOle:J21, GasWigHil:J15}. Fully automated tracking of acoustically active whales from their recorded acoustic data involves numerous algorithmic challenges. In particular, there are typically false positive detections due to noise from the environment and the instrument itself. Furthermore, there are missed detections due to signal directionality and signal masking by background noise. It is thus necessary to solve a data association problem for the automated tracking of acoustically active whales across multiple data snapshots. Data association is complicated in scenarios where multiple acoustically active whales and other acoustic sources are present. In this work, we propose a novel data processing chain that automatically detects and tracks odontocetes in 3-D from their echolocation clicks and demonstrate tracking of Cuvier's beaked whales (\textit{Ziphius cavirostris}) near the coast of California.

\subsection{State-of-the-Art}

An established method to detect acoustic signals produced by whales is to first cross-correlate the time series of acoustic measurements provided by a pair of hydrophones and then apply a detection criterion to the peaks of the resulting cross-correlation signal. In addition, detected peaks can be used to extract time-difference-of-arrival (TDOA) measurements from the cross-correlation signal. Although the conventional cross-correlation\cite{Pap:B84} is well suited for signals with a high signal-to-noise ratio (SNR), the generalized cross-correlation (GCC)\cite{KnaCar:J76} is typically used. The GCC performs a weighting of frequency components to suppress noise, improving the detection performance.

If the TDOA measurements are extracted from four hydrophone pairs and the positions of hydrophones are known, in principle, the 3-D location of a single whale can be computed by solving a nonlinear optimization problem\cite{Zim:B11}. Multiple hydrophones typically form rigid volumetric arrays to facilitate deployment and data processing, recording, and storage. For geometric reasons, just location information in bearing can be provided for whales in the far field of the array. Thus, two or more arrays are deployed for 3-D acoustic source localization. 

In particular, in \onlinecite{GasWigHil:J15} and \onlinecite{WigThaBauHil:R18}, a sequence of 3-D locations of echolocating beaked whales are estimated from the recordings of two high-frequency acoustic recording packages (HARPs) that host tetrahedron-shaped hydrophone arrays. For each snapshot of the acoustic signal, potential direction-of-arrivals (DOAs) relative to each HARP are computed from the TDOA measurements\cite{Zim:B11}. After manually selecting the DOAs that are likely generated from the same whale, the 3-D location of the whale is estimated by computing a least squares solution. The 3-D tracks are the sequence of estimated 3-D locations from DOAs that have been manually selected to be from the same whale.

Multitarget tracking (MTT) methods are sequential Bayesian estimation techniques that automatically infer the number and states of multiple objects from sequences of measurements provided by one or multiple sensors, i.e., without the involvement of human operators. In the context of whale tracking, the states of interest can be the 3-D locations of the whales, the TDOA of whales for a particular hydrophone pair \cite{GruNosOle:J21}, or the frequency of narrow-band whale whistles in the spectrogram\cite{GruWhi:J20}. The MTT methods can succeed in tracking scenarios with false positive detections, missed detections, and data association uncertainty between measurements and objects. Traditional MTT methods such as the joint probabilistic data association (JPDA) filter \cite{BarWilTia:B11} and multiple hypothesis tracker (MHT)\cite{Rei:J79}, model measurements and object states as random vectors. Newer methods like the probabilistic hypothesis density (PHD) filter \cite{Mah:J03} and multi-Bernoulli (MB) filter \cite{Mah:B07,SauCoa:J16}, are derived in the formalism of random finite set (RFS)\cite{Mah:B07}. Recently, a graph-based MTT method that is highly scalable in the number of objects, measurements, and sensors has been proposed \cite{MeyKroWilLauHlaBraWin:J18}. This approach uses particle-based computations to perform operations that can not be evaluated in closed form due to nonlinearities in the system model\cite{MeyBraWilHla:J17}. Graph-based MTT methods for localization and tracking from TDOA measurements in 2-D have been introduced in \onlinecite{MeyTesWin:C17,TesMeyBee:20}.

Some MTT methods have been successfully employed for whale tracking. In \onlinecite{Bag:J15}, the MHT is applied to tracking beaked whales in 3-D. Here, potential 3-D locations of beaked whales are preprocessed from TDOA measurements, which are associated across hydrophone pairs based on click characteristics\cite{Bag:J11}. The preprocessed 3-D locations are used as measurements for an MHT that determines the number of beaked whale tracks, performs data association of 3-D locations with tracks, and runs a Kalman filter for each track. The suboptimum computing of 3-D locations is necessary because the MHT is limited to linear measurement models.
Further inherent challenges of the MHT are its computational complexity and memory requirements \cite{KroMeyCorCarMenWil:J21}. In addition, a Gaussian mixture probability hypothesis density (GM-PHD) filter \cite{VoMa:J06} for the tracking of whales in the TDOA domain is introduced in \onlinecite{GruNosOle:J21}. Both echolocation clicks and whistles of false killer whales are exploited to compute TDOA measurements for multiobject tracking. The data were acquired during line-transect surveys with towed hydrophone arrays. The introduced method extends the original GM-PHD filter by incorporating amplitude information to support the initialization of new whale tracks. However, it relies on linear measurement models and is thus limited to tracking in the TDOA domain.  

Various methods for the localization and tracking of whales from their acoustic signals have been developed. A common approach for the 3-D localization of whales is a grid-search\cite{TiePorFra:J04,Nos:J13,GasWigHil:J15,MacGorGilMalNor:J17,BarNosOle:J21}. wherein an ambiguity surface is generated on a 3-D grid of potential whale locations by comparing the expected modeled TDOA measurements with the actual measurements. In \onlinecite{Nos:J13} and \onlinecite{MacGorGilMalNor:J17}, multiple whales are localized by determining local maxima of the ambiguity surface using the Simplex\cite{NelRog:J65} or the Metropolis-Hastings\cite{MetRosRosTelTel:J53} algorithm. The method in \onlinecite{Nos:J13} iteratively finds whale locations in 3-D by selecting the maximum peak that corresponds to the best match of the TDOA measurements, removing the TDOA measurements corresponding to this maximum peak, and selecting the maximum peak that corresponds to the best match of the remaining TDOA measurements. In \onlinecite{MacGorGilMalNor:J17}, Kalman filtering is used to estimate whale tracks. In the case of simultaneously present whales, multiple Kalman filters run in parallel, and the Hungarian algorithm\cite{Kuh:J55} is used for associating local maxima of the ambiguity surface to Kalman filters. Alternative approaches perform 3-D localization of whales by first estimating the DOAs of acoustic signals from hydrophone arrays\cite{BarGriKliHar:J18,WigThaBauHil:R18,SnyWigBauHil:J19}. The method in \onlinecite{BarGriKliHar:J18} subsequently tracks individual whales using nonsequential estimation based on Gibbs sampling\cite{GemGem:J84}. 

Common challenges of TDOA-based localizing and tracking are finding the correct combination of measurements across hydrophone pairs that correspond to the same acoustic source and initializing whale tracks accordingly. Typically, there are multiple TDOA measurements per hydrophone pair due to noise, echoes, and simultaneous vocalization of multiple whales. This challenge is often referred to as multisensor data association \cite{MeyBraWilHla:J17,MeyWil:J21}. To address this, existing techniques either employ human operators to select and combine measurements manually \cite{GasWigHil:J15}, rely on the local maxima corresponding to incorrectly matched TDOA measurements being significantly lower than those corresponding to true whale locations\cite{Nos:J13}, compute potential whale locations in a brute-force manner, i.e., based on all possible combinations of TDOA measurements \cite{MacGorGilMalNor:J17}, or weigh grid points based on the number of TDOA measurements that are consistent with the corresponding potential whale location \cite{Bag:J11}. All existing methods for TDOA-based localizing and tracking of whales in 3-D either rely on human operators or heuristics to combine TDOA measurements and initialize whale tracks.

\subsection{Contributions, Paper Organization, and Notation}

The fundamental problem addressed in this paper is establishing an algorithmic solution for the tracking of echolocating odontocetes in 3-D. In particular, we aim to develop a data processing method that fully automatically determines the number of odontocetes in the environment and estimates their 3-D tracks from acoustic measurements.

We advocate that this method will make it possible to (i) study deep-diving echolocating odontocetes more objectively and efficiently compared to approaches that rely on human operators and to (ii) reveal key insights on behaviors of odontocetes underwater that otherwise would remain unexplored.

The proposed data processing chain extracts TDOA measurements of echolocation clicks from the raw acoustic signals using a GCC. We propose a variant of the GCC referred to as Whitening Instrument Noise (GCC-WIN) algorithm. This technique aims at suppressing the instrument noise that interferes with the echolocation clicks. The peaks of the TDOAs above a certain amplitude threshold are used to estimate the parameters of interest (i.e., locations and velocities of the whale in time). Odontocetes are first tracked in the TDOA domain using an MTT method based on the framework of factor graphs and the sum-product algorithm (SPA). A second MTT stage estimates odontocetes tracks in 3-D by consistently combining (``fusing'') estimated TDOAs of all hydrophone pairs provided by the first stage. The increased detection rate of the GCC-WIN algorithm results in a lower probability of missing a whale call and, in turn, improved tracking performance. The first tracking stage aims at rejecting false positive TDOA measurements and at resolving longer gaps of missing TDOAs. This first MTT stage significantly improves the performance of estimating odontocetes tracks in 3-D as performed by the second MTT stage.

Tracking whales in 3-D from TDOA measurements is further complicated because the underlying measurement model is non-linear, and the state space is high-dimensional. To address this challenge, we employ a SPA that embeds particle flow\cite{DauHua:C07} and actively migrates particles towards high likelihood regions, making it possible to obtain good object detection and tracking performance in high dimensions \cite{ZhaMey:21}. Contrary to existing methods for detecting and tracking whales in 3-D, the proposed signal processing chain systematically reduces instrument noise and uses a statistical model for multisensor data association and initializing whale tracks. This is expected to improve detection and tracking performance, especially in scenarios with low SNR and a significant number of false positive measurements.

This paper establishes a data processing chain that automatically detects and tracks odontocetes from acoustic measurements of their echolocation clicks. The key contributions of this paper are summarized as follows.
\begin{itemize}
\item We devise a GCC-based algorithm that suppresses the instrument noise to increase the detection probability of echolocation clicks.
\item We apply two stages of graph-based MTT to reject false positives, perform data association, determine the number of odontocetes, and estimate odontocete tracks in 3-D. 
\item We demonstrate the capabilities of our automated tracking method based on echolocation clicks of Cuvier's beaked whales measured by two volumetric hydrophone arrays.
\end{itemize}

\emph{Notation:} Random variables are displayed in sans serif and upright fonts; their realizations in serif, italic fonts. 
Vectors and matrices are denoted by bold lowercase and uppercase letters, respectively. For example, a random variable and its realization are denoted by $\rv x$ and $x$, respectively, and a random vector and its realization by $\RV x$ and $\V x$, respectively. 
Furthermore, $\|\V{x}\|$ and ${\V{x}}^{\text T}$ denote the Euclidean norm and the transpose of vector $\V x$, respectively; 
$\propto$ indicates equality up to a normalization factor;
$f(\V x)$ denotes the pdf of random vector $\RV x$ (this is a short notation for  $f_{\RV x}(\V x)$); 
$f(\V x | \V y)$ denotes the conditional pdf of random vector $\RV x$ conditioned on random vector  $\RV y$  (this is a short notation for  $f_{\RV x | \RV y}(\V x | \V y)$); 
$|\mathcal S|$ denotes the cardinality of set $\mathcal  S$; 
$\M I_n$ denotes the $n \rmv\times\rmv n$ identity \vspace{0mm} matrix. The operator $^{*}$ denotes the complex conjugate, and $\mathpzc{j}=\sqrt{-1}$ is the imaginary unit. Finally, $\RV x_{0:k}$ is short for $[\RV x_0, \dots ,\RV x_k]^{\text T}$.
\vspace{2.5mm}

\section{Generalized Cross Correlation\\ and TDOA Measurements}\label{sec:gcc}

TDOA measurements are typically extracted from pairs of receivers for the localization of an uncooperative source. In 3-D space, each TDOA measurement gives rise to a hyperboloid. With more than two receivers, multiple TDOA measurements can be extracted, and all hyperboloids ideally intersect in a single point at the source location (Fig.~\ref{fig:hyperboloids}).

For TDOA measurement extraction, we compute the cross-correlation between the signals from a spatially separated receiver pair ($s_1$, $s_2$). The receiver pair forms a TDOA sensor $s \in \{1,\dots,n_{\mathrm{s}}\}$, where $n_{\mathrm{s}}$ is the number of sensors, i.e., number of pairs of receivers. The two received signals from a remote source in the presence of noise are modeled as
\begin{linenomath}
\begin{align}
\mathpzc{y}_{s_1}(t) &= \mathpzc{x}_{\hspace{.45mm} s_1}(t) + \mathpzc{n}_{\hspace{.2mm} s_1}(t), \nn\\[.6mm]
\mathpzc{y}_{s_2}(t) &= \alpha \mathpzc{x}_{\hspace{.45mm} s_1}(t + d) + \mathpzc{n}_{s_2}(t), \label{eq:sigModel}
\end{align}
\end{linenomath}
with $\mathpzc{x}_{\hspace{.45mm} s_1}(t)$, $\mathpzc{n}_{s_1}(t)$, and $\mathpzc{n}_{s_2}(t)$ being real, stationary, and ergodic random processes, $\alpha$ being a scaling factor, and $d$ being the TDOA. For an observation interval, $T_{G}$, and a TDOA sensor, $s$, an estimate of the cross-correlation as a function of time delay $\tau$, can be obtained\vspace{2mm} as
\begin{linenomath}
\begin{equation*}
\mathpzc{r}_{\ist s}(\tau) = \frac{1}{T_{G}-\tau}\int_{\tau}^{T_{G}} \rrmv \mathpzc{y}_{\hspace{.15mm} s_1}(t) \ist \mathpzc{y}_{s_2}(t-\tau) \mathrm{d} t.
\vspace{2mm} 
\end{equation*}
\end{linenomath}

An estimate of the cross power spectral density (PSD), $R_{\ist s}(f)$, between the two signals is computed by taking the Fourier transform of the cross-correlation\vspace{.5mm}, i.e.,
\begin{linenomath}
\begin{align*}
R_{\ist s}(f) &= \int_{\infty}^{-\infty} \rmv\rmv \mathpzc{r}_{\hspace{.2mm} s} (\tau) \ist e^{- \mathpzc{j} 2\pi f \tau} \ist \mathrm{d} \tau \nn \\[1.3mm]
           &= Y_{s_1}(f) \ist Y_{s_2}^{*}\rmv(f),
\end{align*}
\end{linenomath}
where $Y_{s_1}(f)$ is the Fourier transform of $\mathpzc{y}_{s_1}(t)$ for the observation interval $T_{G}$. 

The GCC \cite{KnaCar:J76} is defined as the inverse Fourier Transform of the frequency weighted cross PSD, i.e.,
\begin{equation}
   \hat{\mathpzc{r}}_{\ist s}(\tau) = \frac{1}{2\pi} \int_{\infty}^{-\infty} \rmv\rmv \psi(f) \ist R_{\hspace{.2mm} s}(f) \ist e^{\ist \mathpzc{j} 2\pi f t} \, \mathrm{d} f,
   \label{eq:gcc}
\end{equation}
where $\psi(f)$  is the frequency weighting factor. Note that the GCC with $\psi(f)=H_{s_1}(f) H_{s_2}(f)$ can be interpreted as applying linear filters with frequency responses $H_{s_1}(f)$ and $H_{s_2}(f)$ to the signals $\mathpzc{y}_{s_1}(t)$ and $\mathpzc{y}_{s_2}(t)$, respectively, and subsequently performing a conventional cross-correlation.

Frequency weighting is performed according to specific optimization criteria. In applications where the noise PSD is unknown, popular choices of frequency weighting factors result in the smoothed coherence transform (SCOT) \cite{CarNutCab:J73} and phase transform (PHAT) \cite{KnaCar:J76}. The SCOT normalizes the PSDs of the individual signals to unit magnitude, i.e., $\psi_{\text{SCOT}}(f) = 1/( R_{s_1,s_1}(f) R_{s_2,s_2}(f) )^{1/2}$, where $R_{s_1,s_1}$ denotes the auto PSD of $\mathpzc{y}_{s_1}(t)$. Similarly, the PHAT normalizes the cross PSD to unit magnitude, i.e., $\psi_{\text{PHAT}}(f) =1/|R_{s}(f)|$. 

In this work, it is assumed that the auto PSDs $G_{s_1,s_1}(f)$ and $G_{s_2,s_2}(f)$ of the noise $\mathpzc{n}_{s_1}(t)$ and $\mathpzc{n}_{s_2}(t)$ are mainly dominated by instrument noise. It can be measured or precomputed and is thus considered known. The optimal frequency weighting for the whitening instrument noise (WIN) is given by $\psi_{\text{WIN}}(f) =1/( G_{s_1,s_1}(f) \ist G_{s_2,s_2}(f) )^{1/2}$. The resulting GCC Eq.~\ref{eq:gcc} can be interpreted as applying the noise whitening filters $H_{s_1}(f) = 1/(G_{s_1,s_1}(f))^{1/2}$ and $H_{s_2}(f) = 1/(G_{s_2,s_2}(f))^{1/2}$ to the signals $y_{s_1}(t) $ and $y_{s_2}(t)$, respectively, and subsequently performing a conventional cross-correlation. A detailed discussion of the resulting GCC-WIN technique will be presented in Section \ref{sec:methods}.

To enhance the probability of detection, TDOA measurements can be either (i) computed based on a sequence of click trains \cite{Nos:J13, Bag:J15} or (ii) obtained by extracting TDOAs of individual echolocation clicks and combining them into a single measurement. In the presence of highly-correlated noise, the former approach is unsuitable. Thus, we consider the latter approach. It is assumed that the whale is stationary over a time interval, $T_M$, that is longer than $T_G$, i.e., $T_M > T_G$. ($T_M$ is the duration of the time between the discrete-time step $k$ of the considered tracking algorithms.) For each sensor, TDOAs of individual echolocation clicks are computed by finding the peaks of the cross PSDs that are above a certain threshold $P_{\mathrm{TDOA}}$. The peaks of multiple observation intervals of length $T_G$ are then accumulated over a time interval of duration $T_M$. The\vspace{-.5mm} resulting set of TDOAs $\V{z}^{(m)}_{k,s}$, $m \rmv\in\rmv \{1,\dots,m_{k,s}\}$ is considered as TDOA measurements of clicks generated by odontocetes. The TDOA measurements of all sensors are used as input for multitarget tracking.

\section{Multi-target Tracking}\label{sec:}

A key challenge of multitarget tracking (MTT) from sequences of measurements provided by one or multiple sensors is that the origin of measurements is typically unknown, i.e., it is not clear which target originated which TDOA measurement. Moreover, since the number of targets is also unknown, it has to be estimated directly from the data. \vspace{2mm}
\subsubsection{MTT With Perfect Measurement-to-Object Associations} Assuming that the origin of each measurement is perfectly known, i.e., measurement-to-object associations are either provided by a perfect human operator or a data association algorithm, the MTT problem can be split up into multiple parallel single target tracking problems. Here, a sequential Bayesian estimation or Bayes filter \cite{ShaKirLi:B02,AruMasGorCla:02} is typically employed to estimate the state of each target individually and recursively. Define the target state and its associated measurements for all $n_{\mathrm{s}}$ sensors and at a discrete time step $k$, as random vectors $\RV x_k$ and $\RV z_k = [\RV z^{\text T}_{k,1}, \dots, \RV z^{\text T}_{k,n_{\mathrm{s}}}]^{\text T}$, respectively. 

The state $\RV x_k$ typically consists of the target's position and motion-related parameters. We are interested in estimating the state from the available measurements up to time $k$, $\RV z_{1:k}$. Given the conditional probability density function (pdf) of the state given the measurements, $f(\V x_k|\V z_{1:k})$, the minimum mean-squared error (MMSE) estimate of the state of a single target, $\V x_k$, can be found as\vspace{-.5mm} \cite{Kay:B93}
\begin{equation}
   \hat{\V x}_{k}^{\mathrm{MMSE}} = \int\;\V x_k f(\V x_k|\V z_{1:k}) d\V x_{k}.
   \label{eq:MMSE}
   \vspace{-1mm}
\end{equation}

To obtain $f(\V x_k|\V z_{1:k})$, one could na\"ively marginalize the available joint pdf $f(\V x_{1:k}|\V z_{1:k})$. This approach, however, suffers from the curse of dimensionality since the dimension of $\RV x_{1:k}$ grows with each time step, and, as a result, the computational complexity of na\"ive marginalization increases exponentially and becomes intractable. The Bayes filter exploits that a first-order Markov process can describe the statistical model of single object tracking to reduce computations. At each time $k$, a prediction and $n_{\mathrm{s}}$ update steps \cite{ShaKirLi:B02,AruMasGorCla:02} are performed, and the resulting sequential processing schemes yields a computational complexity that is linear with time $k$. MTT methods are sequential Bayesian estimation methods that also consider measurement-origin uncertainty (MOU) and the unknown number of states to be estimated.
\vspace{2mm}

\subsubsection{MTT With MOU and Known Number of Targets} Consider an MTT problem with multiple sensors where the number of targets is known, but measurements are subject to MOU. In addition, there are false positives, i.e., measurements that have not been generated by any target, and missed detections, i.e., present targets, may not generate a measurement. It is assumed that there are $i \in \{1,\dots,n_\mathrm{t}\}$ targets. At time $k$, the state of the $i$\textsuperscript{th} target is denoted as $\RV x_k^{(i)}$. For future reference, we introduce the notation $\RV x_k = \big[\RV x_k^{(1)}, \dots ,\RV x_k^{(n_\mathrm{t})} \big]^{\text T}$. Each target state evolves independently according to the Markovian state transition pdf, i.e., $f(\V x_k|\V x_{k-1}) = \prod^{n_{\mathrm{t}}}_{i=1} f(\V x_k^{(i)}|\V x_{k-1}^{(i)})$.

Each sensor $s \in \{1,\dots,n_s\}$ produces $m_{k,s}$ TDOA measurements $\RV{z}_{k,s} = \big[\RV z_{k,s}^{(1)}, \dots ,\RV z_{k,s}^{(m_{k,s})} \big]^{\text T}\rmv\rmv$. It is assumed that each measurement either originates from the target or is a false positive and that a target generates at most one measurement at each sensor. Measurement generation of target $i$ at sensor $s$ is modeled by a Bernoulli experiment characterized by the probability of detection $p_d^{(s)}(\V x_k^{(i)})$. If the target with state $\V x_k^{(i)}$ generates a measurement $\RV z_{k,s}^{(m)}$, the measurement is distributed according to  $f(\V z_{k,s}^{(m)} | \V x_k^{(i)})$. The number of false\vspace{0mm} positives at each time step is Poisson distributed with a mean $\mu_{\mathrm{fp}}^{(s)}$. False positives are independent of the measurements that have originated from the targets. They are also independent and identically distributed (iid) according to pdf $f_{\mathrm{fp}}^{(s)}(\V z_{k,s}^{(m)})$. 

At time $k$ and sensor $s$, the unknown association between measurements and targets is modeled by the latent random vector $\RV{a}_{k,s} = \big[\rv a_{k,s}^{(1)}, \dots ,\rv a_{k,s}^{(n_\mathrm{t})} \big]^{\text T}$ that is composed of random variables, $\rv a_{k,s}^{(i)}$, defined as \cite{BarWilTia:B11}
\vspace{.5mm}
\begin{equation}
   \rv{a}_{k,s}^{(i)} = 
      \begin{cases}
         m \in \{1,...,m_{k,s}\} & \text{\parbox[t]{7cm}{at time $k$ and sensor $s$, target $i$ generates measurement $m$}}\\
         \\[-3mm]
         0                       & \text{otherwise.}
      \end{cases}
\end{equation} 
For future reference, we introduce the joint vectors $\RV{a}_{k} = \big[\rv a_{k,1}^{(1)}, \dots ,\rv a_{k,n_{\mathrm{s}}}^{(n_{\mathrm{t}})} \big]^{\text T}$ and $\RV{z}_{k} = \big[\RV z_{k,1}, \dots ,\RV z_{k,n_{\mathrm{s}}} \big]^{\text T}$.

The restriction that there can be at most one measurement associated with a target at every time step can be checked by the following indicator\vspace{.5mm} function
\begin{linenomath}
\begin{equation}
   \psi(\V a_{k,s}) = 
   \begin{cases}
      0 & \text{\parbox[t]{7cm}{$\exists i,i' \in \{1,\dots,n_\mathrm{t}\}$ such that $i \neq i'$ and $a_{k,s}^{(i)} = a_{k,s}^{(i')} \neq 0$}} \\ 
      \\[-3mm]
      1 & \text{otherwise.}\label{eq:indication}
   \end{cases}
   \vspace{.1mm}
\end{equation} 
\end{linenomath}

The marginal posterior pdf, $f(\V x_k^{(i)}|\V z_{1:k})$, used for state estimation according to  Eq.~\ref{eq:MMSE} could again be found by marginalization, i.e., $f(\V x_k^{(i)}|\V z_{1:k}) = \int \sum_{\V a_{1:k}} f(\V x_{0:k},\V a_{1:k}|\V z_{1:k}) \; \mathrm{d} \V{x}_{\sim k}^{(i)}$ where the vector $ \V{x}_{\sim k}^{(i)}$ is equal to $\V{x}_{1:k}$ with state $\V x_k^{(i)}$ removed. However, the computational complexity of this na\"ive marginalization would again be infeasible due to the reasons discussed above. To reduce computational complexity, one can again exploit the fact that the posterior pdf $ f(\V x_{0:k}, \V a_{1:k}|\V z_{1:k}) $ follows a first-order Markov process, i.e.,
\begin{linenomath}
\begin{align}
   &f(\V x_{0:k},\V a_{1:k}|\V z_{1:k}) \notag \\[.5mm]
   &\hspace{5mm}\propto f(\V x_{0}) \prod_{k'=1}^{k} f(\V x_{k'}|\V x_{k'-1}) f(\V z_{k'},\V a_{k'} |\V x_{k'}) \nn\\
   &\hspace{5mm}= f(\V x_{0}) \prod_{k'=1}^{k} f(\V x_{k'}|\V x_{k'-1}) \prod^{n_{\mathrm{s}}}_{s=1} f(\V z_{k',s},\V a_{k',s} |\V x_{k'}).\label{eq:mttAKnownNo}
\end{align}
\end{linenomath}
Here, $f(\V x_{k} |\V x_{k-1})$ is the state-transition pdfs discussed above, $f(\V x_{0})$ is an arbitrary prior at time $k=0$, and $f(\V z_{k,s},\V a_{k,s} |\V x_{k})$ is the conditional pdf that models the MOU measurement generation process. In the last line of Eq.~\ref{eq:mttAKnownNo}, we have used the fact that conditioned on the target states; measurement generation is independent across sensors. Note that $f(\V z_{k,s},\V a_{k,s} |\V x_{k})$ is a function of $p_d^{(s)}(\V x_k^{(i)})$, $\mu_{\mathrm{fp}}^{(s)}$, $f_{\mathrm{fp}}^{(s)}(\V z_{k,s}^{(m)})$, and $f(\V z_{k,s}^{(m)} | \V x_k^{(i)})$ (see \onlinecite{BarWilTia:B11} for details). Based on the factorization of the statistical model in Eq.~\ref{eq:mttAKnownNo}, e.g., a sequential Bayesian estimation approach referred to as probabilistic data association filter \cite{BarWilTia:B11} can be developed.
 
\subsubsection{MTT With MOU and Unknown Number of Targets}\label{sec:mttUnknown} In real-world scenarios, such as the whale tracking problem,  the number of targets is time-varying and unknown. To account for this, potential target (PT) states can be introduced \cite{MeyKroWilLauHlaBraWin:J18}. Suppose that at time $k$, we have PTs with indexes $j \in \{1,\dots,j_k\}$. A binary variable, $r_k^{(j)} \in \{0,1\}$, indicates the existence of the PT $j$, where $r_k^{(j)} \rmv=\rmv 1$ if and only if PT $j$ exists. The augmented state of PT $j$ is given by $\RV y_{k}^{(j)} = [\RV x_{k}^{(j){\text T}} \ist\ist r_k^{(j)}]^{\text T}$ where $\RV{x}_{k}^{(j)} $ consists of the target's position and further motion-related parameters. There are two types of PTs:
\begin{itemize}
    \item \textit{New PTs} represent targets that, for the first time, have generated a measurement. Their states are denoted by $\overline{\V{y}}{}^{(j)}_{k,s} = [\overline{\V{x}}{}_{k,s}^{(j) \ist \text T} \hspace{1mm} \overline{r}{}_{k,s}^{(j)}]^{\text T}$. At time $k$ and sensor $s$, a new PT is introduced for each measurement $m \in \{1, \dots, m_{k,s}\}$.
    \vspace{.8mm}
    
    \item \textit{Legacy PTs} represent targets that already have generated at least one measurement at a previous time step $k^\prime < k$ or previous sensor $s^\prime < s$. Their states are denoted by $\underline{\V{y}}^{(j)}_{k,s} = [\underline{\V{x}}_{k,s}^{(j) \text T} \hspace{1mm} \underline{r}^{(j)}_{k,s}]^{\text T}$.
    \vspace{1mm}
\end{itemize}
We denote by $\underline{\V{y}}_k$ and $\overline{\V{y}}_k$ the\vspace{.2mm} vectors that consist of all legacy and new PT states, respectively, and by\vspace{.2mm} $\V{y}_k = [\underline{\V{y}}_k \ist \overline{\V{y}}_k ]^{\text T}$ the vector that consists of all PT state at time $k$. 

At each time $k$, the number of targets that for the first time have generated a measurement at sensors $s$ are Poisson distributed with mean $\mu_n^{(s)}$. The states of these newly detected targets are iid with pdf $f^{(s)}_n(\V x_{k}^{(j)})$. Newly detected targets are statistically independent of existing targets.  A PT $j$ that existed at time $k\rmv-\rmv1$ continues to exist at time $k$ with survival probability $p_{\mathrm{su}}(\V x_{k}^{(j)})$. All PTs at time $k\rmv-\rmv1$ become legacy PTs at time $k$.
\vspace{2mm}

To reduce computational complexity, one can again exploit structure in the factorization of the posterior pdf $ f(\V y_{1:k},\V a_{1:k}|\V z_{1:k}) $. In particular, using common Markov assumptions \cite{MeyKroWilLauHlaBraWin:J18}, $ f(\V y_{1:k},\V a_{1:k}|\V z_{1:k}) $ factorizes according\vspace{-.2mm} to
\begin{linenomath}
\begin{align}
   f(\V y_{1:k},\V a_{1:k}|\V z_{1:k})  &\propto f(\V z_{1:k},\V a_{1:k},\V y_{1:k})\notag\\[.3mm]
   &= f(\V y_{0})\prod_{k'=1}^{k} f(\V z_{k'},\V a_{k'},\V y_{k'}|\V y_{k'-1}).\label{eq:markovAssumption}
\end{align}
\end{linenomath}
Upon explicitly distinguishing between the legacy and new PTs and by exploiting the chain rule for pdfs, the conditional pdf of current measurements, current association variables, and current states given the previous states can be expanded\vspace{-4mm} as
\begin{linenomath}
\begin{align}
   & f(\V z_k,\V a_k,\V y_{k}|\V y_{k-1}) \nn\\[.7mm]
   & \hspace{7mm}= f(\V z_k,\V a_k, \underline{\V y}_{k},\overline{\V y}_{k}|\V y_{k-1})\notag\\[.7mm]
      & \hspace{7mm}= f(\underline{\V y}_{k}|\V y_{k-1}) f(\V z_k,\V a_k, \overline{\V y}_{k}|\underline{\V y}_{k},\V y_{k-1}). \label{eq:eq1}
\end{align}
\end{linenomath}
 Now, one can use the fact that given the legacy PT states at time $k$, (i) the new PT states at time $k$ are conditionally independent of the previous PT states at time $k-1$ and that (ii) measurements, association variables, and new PTs are statistically independent across the $n_{\mathrm{s}}$ sensors, i.e.,
\begin{linenomath}
\begin{align}
  &f(\V z_k,\V a_k,\V y_{k}|\V y_{k-1}) \nn\\[1mm]
  &\hspace{6mm}= f(\underline{\V y}_{k}|\V y_{k-1}) f\big(\V z_{k},\V a_{k},\overline{\V y}_{k}|\underline{\V y}_{k} \big).\nn\\[1mm]
  &\hspace{6mm}= f(\underline{\V y}_{k}|\V y_{k-1}) \prod^{n_{\mathrm{s}}}_{s = 1} f\big(\V z_{k,s},\V a_{k,s},\overline{\V y}_{k,s}|\underline{\V y}_{k,s-1} \big), \label{eq:eq2}
\end{align}
\end{linenomath}
where $\underline{\V y}_{k,0} = \underline{\V y}_{k}$ and $\underline{\V y}_{k,s} = \big[\underline{\V y}^{\text T}_{k,0}, \overline{\V y}^{\text T}_{k,1}, \dots, \overline{\V y}^{\text T}_{k,s-1} \big]^{\text T}$.

In addition, targets move in time independently, and therefore, the state-transition pdf of the legacy PTs reads
\begin{equation}
   f(\underline{\V y}_{k}|\V y_{k-1}) = \prod_{j=1}^{j_{k-1}} f(\underline{\V y}_{k}^{(j)}|\V y_{k-1}^{(j)}).\label{eq:augstatetransition}
\end{equation}
The state-transition model $f(\underline{\V y}_{k}^{(j)}|\V y_{k-1}^{(j)})$ is a function of the survival probability $p_{\mathrm{su}}(\V x_{k}^{(j)})$.

By plugging Eq.~\ref{eq:augstatetransition} into Eq.~\ref{eq:eq2} and, in turn, Eq.~\ref{eq:eq2} into Eq.~\ref{eq:markovAssumption}, and by making use of the functional form of $ f\big(\V z_{k,s},\V a_{k,s},\overline{\V y}_{k,s}|\underline{\V y}_{k,s-1} \big) $ (see \onlinecite{MeyKroWilLauHlaBraWin:J18} for details), the joint posterior pdf of $\RV y_{1:k}$ and $\RV a_{1:k}$ given $\RV z_{1:k}$ becomes
\begin{linenomath}
\begin{align}
  & f(\V y_{1:k},\V a_{1:k}|\V z_{1:k}) \notag\\
      &\hspace{5mm}\propto \prod_{k'=1}^{k} \Big( \prod_{j'=1}^{j_{k'-1}} f(\underline{\V y}_{k'}^{(j')}|\underline{\V y}_{k'-1}^{(j')})\Big) \ist \prod_{s=1}^{n_s} \ist \psi(\V a_{k',s}) \notag\\
      &\hspace{5mm}\times \Big( \prod_{j=1}^{j_{k',s}} q(\underline{\V x}_{k',s}^{(j)}, \underline{r}_{k',s}^{(j)}, a_{k',s}^{(j)}; \V z_{k',s}) \Big) \notag\\
      &\hspace{5mm}\times \prod_{m=1}^{m_{k',s}} v(\overline{\V x}_{k',s}^{(m)}, \overline{r}_{k',s}^{(m)}, a_{k',s}^{m}).\label{eq:finalMTT}
\end{align}
\end{linenomath}
Here, the factors $q(\underline{\V x}_{k,s}^{(j)}, \underline{r}_{k,s}^{(j)}, a_{k,s}^{(j)}; \V z_{k,s})$ and $v(\overline{\V x}_{k,s}^{(m)},\overline{r}_{k,s}^{(m)},$ $a_{k,s}^{m})$ are functions of $p_d^{(s)}(\V x_k^{(i)})$, $\mu_{\mathrm{fp}}^{(s)}$, $f_{\mathrm{fp}}^{(s)}(\V z_{k,s}^{(m)})$, and $f(\V z_{k,s}^{(m)} | \V x_k^{(i)})$ (see \onlinecite{MeyKroWilLauHlaBraWin:J18} for details).

For target detection and estimation, the marginal pdfs $f(\V x^{(j)}_k\rmv\rmv, r^{(j)}_k|\V z_{1:k}) \rmv=\rmv f(\V y^{(j)}_k|\V z_{1:k})$ are required. In particular, target detection is performed by introducing a threshold $P_{\text{th}}$ that is compared with the existence probability $p\big(r_{k}^{(j)}\! \rmv= 1 \big| \V{z}_{1:k} \big)$, i.e., PT $j \in \{1,\dots,j_k\}$ is declared to exist if $p\big(r_{k}^{(j)}\! \rmv= 1 \big| \V{z}_{1:k} \big) \rmv>\rmv P_{\text{th}}$. Note that $p\big(r_{k}^{(j)}\! \rmv=\rmv 1 \big| \V{z}_{1:k} \big) \rmv= \int f\big(\V{x}_k^{(j)}, r_k^{(j)}\! \rmv= 1 \big| \V{z}_{1:k}\big) \ist\mathrm{d}\V{x}_k^{(j)}$.
For PTs declared to exist, state estimation is performed by computing the minimum-mean-squared error estimate \cite{Kay:B93} as
\vspace{.1mm}
\begin{equation}
\hat{\V{x}}_k^{(j)}
\ist\triangleq \int \V{x}_k^{(j)} f\big(\V{x}_k^{(j)} \big| r_k^{(j)} \rmv=\rmv 1, \V{z}_{1:k}\big) \ist\mathrm{d}\V{x}_k^{(j)},
\label{eq:mmseEst}
\end{equation} 
where $f\big(\V{x}_k^{(j)} \big| r_k^{(j)} \rmv=\rmv 1, \V{z}_{1:k}\big) \rmv=\rmv f\big(\V{x}_k^{(j)}, r_k^{(j)} \rmv=\rmv 1 \big| \V{z}_{1:k}\big)/ p\big(r_{k}^{(j)}$ $=\rmv 1\big| \V{z}_{1:k} \big)$. For direct computation of $f(\V y_k|\V z_{1:k})$, one could again na\"ively marginalize the available joint pdf $f(\V y_{1:k},\V a_{1:k}|\V z_{1:k})$. This marginalization can be performed efficiently by the framework of factor graphs and the sum-product algorithm (SPA) \cite{KscFreLoe:01}. The complete system model and the SPA for MTT can be found in \onlinecite{MeyKroWilLauHlaBraWin:J18}.

The tracking of objects in 3-D from TDOA measurements is further complicated because the underlying measurement model is non-linear, and the state space is high-dimensional. To address this challenge, we employ a SPA that embeds particle flow. Here, to perform the SPA effectively, particles are migrated towards regions of high likelihood based on the solution of a partial differential equation. This makes it possible to obtain good object detection, and tracking performance in 3-D \cite{ZhaMey:21}.

\section{The Proposed Data Processing Chain}

\begin{figure}
   \includegraphics[width=\linewidth]{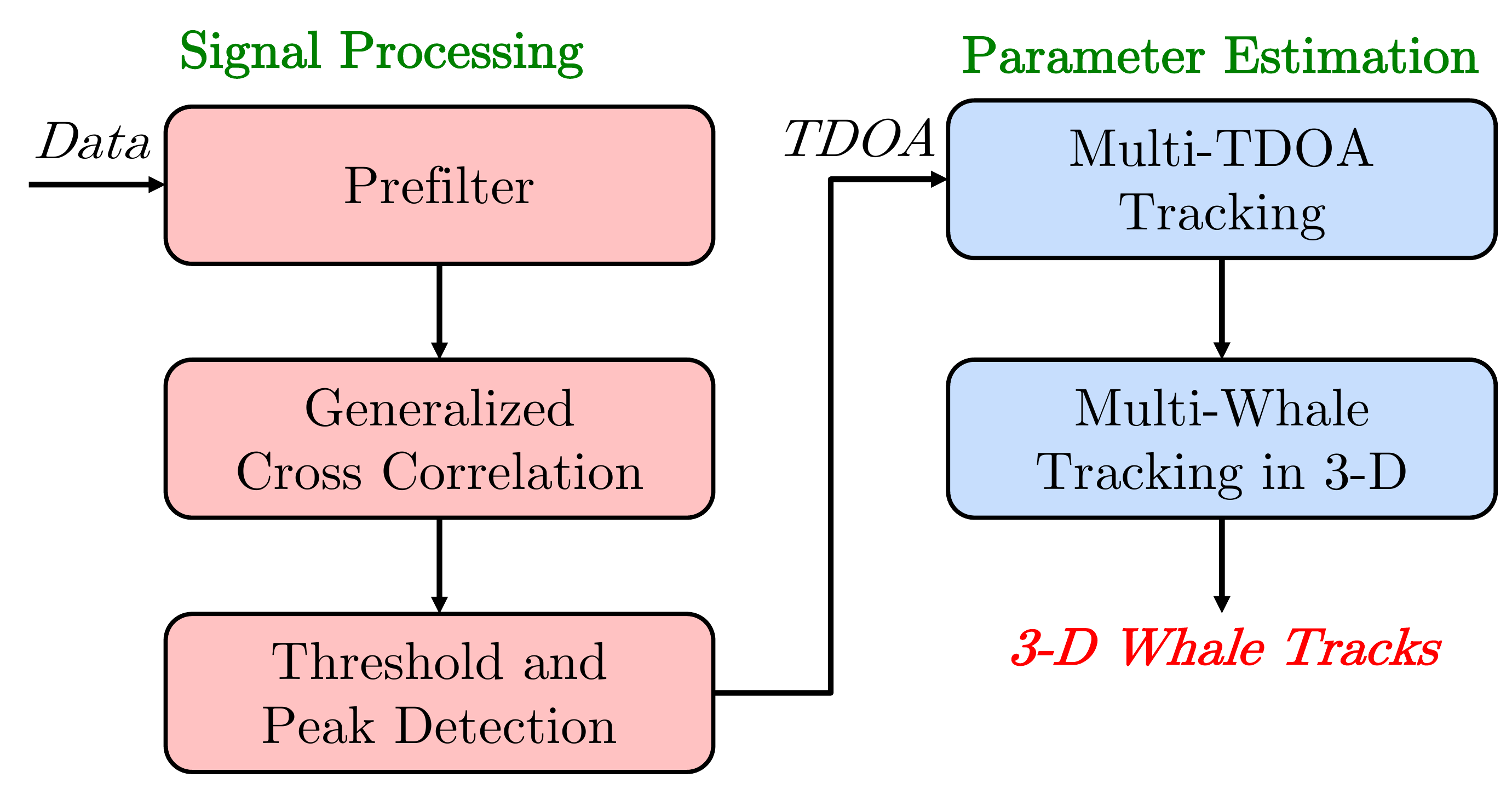}
   \caption{Data processing chain for the detection and tracking of odontocetes in 3-D from their echolocation clicks. TDOAs of echolocation clicks are computed in the signal processing module, which performs TDOA extraction. 3-D whale tracks are computed in the parameters estimation module which performs sequential Bayesian estimation for MTT. \vspace{-4mm}}
   \label{fig:overview}
\end{figure}

The proposed data processing chain performs two main tasks: (1) signal processing and (2) parameter estimation (Fig.~\ref{fig:overview}). In the signal processing module, pre-filters are first applied to the raw acoustic signal. Then, the GCC is performed to extract time delay peaks with GCC amplitudes above a threshold, $P_{TDOA}$.

\label{sec:methods}

In the parameter estimation module of the processing chain, we are interested in estimating the locations and velocities of the echolocating odontocetes. Odontocetes are first tracked in the TDOA domain and then in 3-D. In both cases, we apply the tracking algorithm based on the SPA described in Sec. \ref{sec:mttUnknown}. In the TDOA domain, both the motion and the measurement models are linear. Tracking first in the TDOA domain makes it possible to reduce the number of false positives significantly. A low number of false positives is essential for successful tracking in 3-D. Finally, the output of the TDOA tracker is used as input for whale tracking in 3-D.

\subsection{Echolocation Click Detection and TDOA estimates}\label{sec:gccImpl}

The echolocation clicks are characterized by their short pulse length and broad bandwidth\cite{BauMcDSimSolMerOleRocWigRanYacHil:J13}. Depending on prior knowledge of the noise, different adaptions of GCC can be utilized to detect the echolocation clicks and estimate the TDOA. In this work, we are interested in maximizing the number of detected clicks to reduce the duration and frequency of data gaps in time that can significantly hinder the performance of the tracking algorithms. The echolocation clicks are challenging to detect if their SNR is low. The SNR depends on various factors, such as the distance between the source and the hydrophones, the animal orientation and its echolocation click beam direction with respect to the receiver, as well as ambient or system noise.

Each GCC of length $T_G$ corresponds to $N_G$ samples of the discrete time signal and eventually results in a discrete tracking time step, $n$. Following Eq.~\ref{eq:sigModel}, the discrete received signals for sensor $s$, at time $n$\vspace{-1mm} are given by
\begin{linenomath}
\begin{align*}
   \mathpzc{y}_{s_1}[n] &= \mathpzc{x}_{s_1}[n] + \mathpzc{n}_{s_1}[n] \nn\\[2mm]
   \mathpzc{y}_{s_2}[n] &= \alpha \mathpzc{x}_{s_1}[n + d] + \mathpzc{n}_{s_2}[n]. \nn\\[0mm]
   \nn\\[-11mm]
\end{align*}
\end{linenomath}
The discrete Fourier transform pairs of $\mathpzc{y}_{s_1}[n]$ and $\mathpzc{y}_{s_2}[n]$ are $Y_{s_1}[l]$ and $Y_{s_2}[l]$, respectively, where $l$ is the discrete frequency. The GCC as a function of discrete time delay, $m$, \vspace{-1.5mm} is
\begin{linenomath}
\begin{equation*}
   \hat{\mathpzc{r}}_{s}[m] = \frac{1}{N_G} \sum_{l'=0}^{N_G-1} \rmv\rmv \psi[l] \ist R_{s}[l] \ist e^{\ist \mathpzc{j} 2\pi m l / N},
   \label{eq:gccDiscrete}
\end{equation*}
\end{linenomath}
where $R_{s}[l] = Y_{s_1}[l]\,Y_{s_1}^{*}[l]$ is the auto PSD estimate of the received signals, and $\psi[l]$ is the frequency weighting of the GCC.

\begin{figure}
   \centering
   \includegraphics[width=\linewidth]{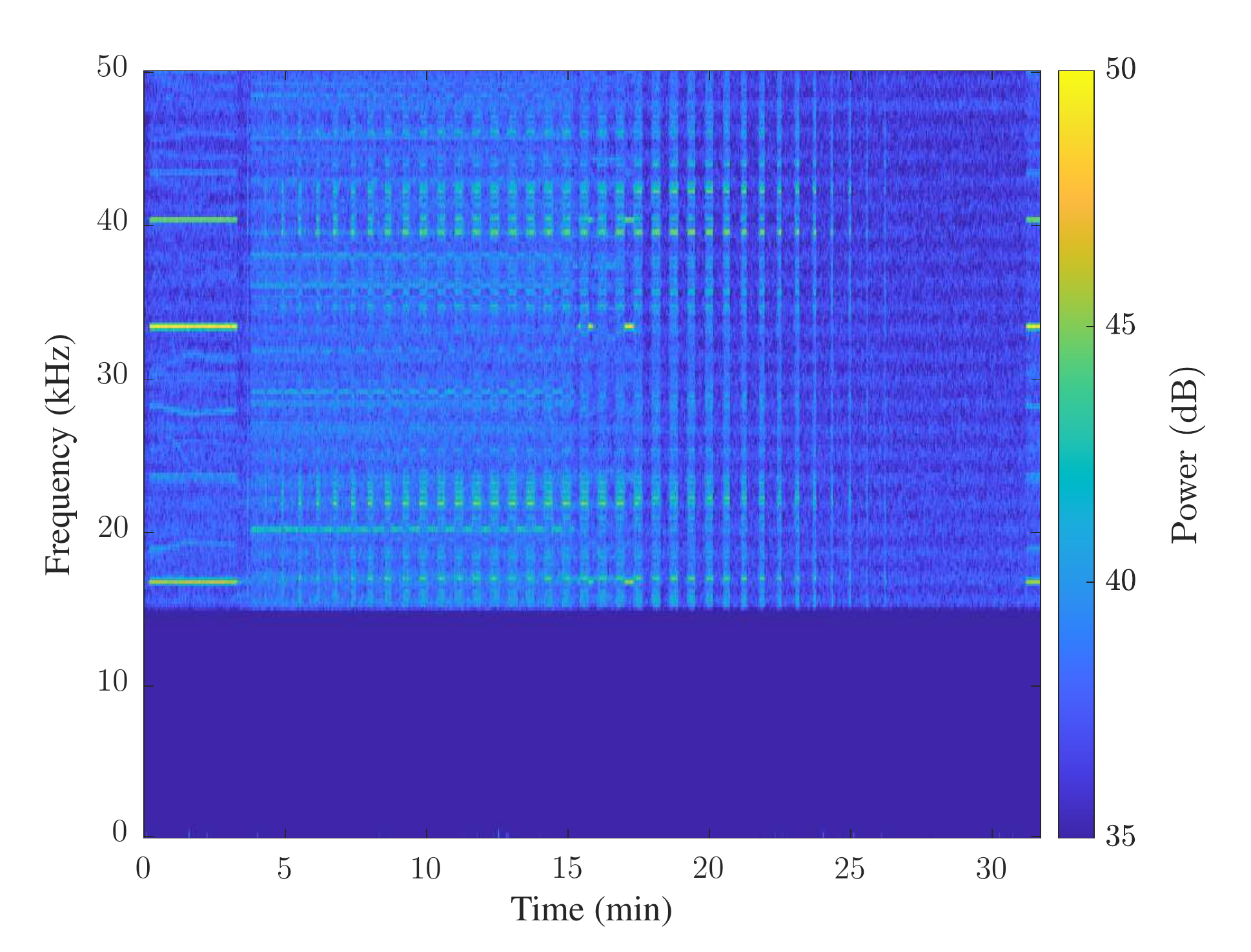}
   \caption{Exemplary spectrogram of the instrument noise. The spectrogram repeats periodically. Each column of the spectrogram serves as an estimated noise PSD for GCC-WIN. The considered data was collected on July 1\textsuperscript{st}, 2018. For each hydrophone, an individual spectrogram is extracted.}
   \label{fig:noiseTemplate}
\end{figure}

In this work, it is assumed that an accurate estimate of the auto PSD of the instrument noise is available and can be used within a GCC-WIN approach. In particular, the cross PSD estimate, $R_{s}[l]$, is normalized by the auto PSD estimates of the known noise (see Sec.~\ref{sec:gcc}). Let $G_{s_1,s_1}[l]$ and $G_{s_2,s_2}[l]$ be the auto PSD estimates of the noise at the respective receivers. The frequency weighting used in GCC-WIN then reads $\psi_{s}[l]=1/(G_{s_1,s_1}[l]\,G_{s_2,s_2}[l])^{1/2}$. In case the statistics of the noise produced by the instrument are time-varying but periodic, as it is the case for the HARP, a sequence of time-varying noise PSD estimates is extracted from precomputed spectrograms of the noise signal, i.e., from portions of signals without echolocation clicks (example in Fig.~\ref{fig:noiseTemplate}). A concrete implementation of this procedure for acoustic data provided by the HARP, is presented in Sec.~\ref{sec:sigProcImpl}.

Finally, as described in Sec.~\ref{sec:gcc}, we accumulate TDOAs extracted from individual echolocation clicks over a longer interval, $T_M$, to increase the probability of detection. Accumulating TDOAs, however, can lead to multiple TDOAs corresponding to one whale per time step $k$ due to noise. Hence, a clustering of measurements is necessary to enforce the assumption that each whale generates at most one measurement. The following clustering procedure led to the best tracking results. First, large clusters are formed by finding TDOAs that are $n_c$ samples apart and grouping. Within each large cluster, TDOAs are further separated into smaller clusters based on the local minima of the GCC-WIN amplitudes. The TDOAs within each small cluster are weighted based on their amplitudes and merged to generate a single TDOA measurement per small cluster.

\subsection{MTT of Echolocating Odontocetes}\label{sec:mttImpl}

In this section, we will provide details on how the MTT framework presented in Section \ref{sec:mttUnknown} is applied to the tracking of odontocetess. In particular, we will discuss the statistical model of the two MTT stages. The first stage operates in the TDOA domain and is applied to each hydrophone pair, i.e., TDOA sensor, separately and in parallel. The second stage operates in a 3-D Cartesian coordinate system and fuses the results provided by the first stage.

\subsubsection{MTT in TDOA Domain} For the tracking in the TDOA domain, at time step $k$ and sensor $s$, the state of odontocetes $j$ is given by $\RV d^{(j)}_{k,s} = [\rv d^{(j)}_{k,s} \; \dot{\rv d}^{(j)}_{k,s}]^{\text T}$, where\vspace{-.3mm} $\rv d^{(j)}_{k,s}$ is the true TDOA of the clicks originated by odontocetes $j$ at sensor $s$, and  $\dot{\rv{d}}^{(j)}_{k,s}$ is its rate of change. A linear constant velocity motion model \cite{BarWilTia:B11} is considered, i.e.\vspace{-1mm},
\begin{equation}
   \RV d_{k,s}^{(j)} = 
   \begin{bmatrix} 
      1 & T_{M} \\ 
      0 & 1
   \end{bmatrix}
   \begin{bmatrix}
      \rv d_{k-1,s}^{(j)} \\ \dot{\rv{d}}_{k-1,s}^{(j)}
   \end{bmatrix}
   + \RV u_{k,s}^{(j)}
\end{equation}
where the driving noise $\RV u_{k,s}^{(j)} \in \mathbb{R}^2$ is a zero-mean multivariate Gaussian random vector with covariance\vspace{2mm} matrix 
\begin{equation}
\M{\Sigma}_{\RV{u}} = 
 \begin{bmatrix}
  \frac{T_{M}^3}{3} & \hspace{-1.5mm}\frac{T_{M}^2}{2} \\[.7mm]
  \frac{T_{M}^2}{2} &  \hspace{-1.5mm}T_{M} \\
 \end{bmatrix} \rmv\rmv\sigma^2_{\RV{u}}
\end{equation}
and driving noise standard deviation (std) $\sigma_{\RV{u}}$. 

The additive-Gaussian measurement model is given as follows. Let the TDOA measurement $\rv{z}^{(m)}_{k,s}$ be originated by the odontocetes with index $j$ at sensor $s$. The measurement model\vspace{-1mm} then reads
\begin{equation}
   \rv{z}^{(m)}_{k,s}= \rv d_{k,s}^{(j)}  + \rv{v}^{(m)}_{k,s}.
\end{equation}
Here, $\rv{v}^{(m)}_{k,s}$ is a zero-mean\vspace{0mm} Gaussian measurement noise with std $\sigma_{\rv{v}}$. The pdf that characterizes false positives,  $f_{\mathrm{fp}}^{(s)}\big(z_{k,s}^{(m)}\big)$, is uniform on the interval $[-T_s^{\ist \text{max}} \hspace{1mm} T_s^{\ist \text{max}}]$, where $T_s^{\ist \text{max}}$ is the maximum time delay that can be measured by sensor $s$, i.e., 
\begin{equation}
T_s^{\ist \text{max}} =  \|\V q_{s_1} \rmv-\rmv \V q_{s_2}\| / c
\end{equation}
where $q_{s_1}\in\mathbb{R}^3$ and $q_{s_2}\in\mathbb{R}^3$ are the positions of the hydrophone pair $(s_1,s_2)$ that defines sensor $s$ and $c$ is the speed of sound.  

In the considered linear MTT problem, there is MOU, and the number of odontocetess is unknown and time-varying. Thus, the SPA-based MTT method described in Section \ref{sec:mttUnknown} is employed. The results of this first stage are sets of TDOA estimates $\hat{d}_{k,s}^{(j)}$, $j \in \{1,\dots, j_{k,s}\}$ for each time step $k$ and each sensor $s$. These TDOA estimates are used as measurements in the 3-D MTT tracking stage. In what follows, we will denote the aforementioned set of TDOA estimates as $\hat{d}_{k,s}^{(m)}$, $j \in \{1,\dots, m_{k,s}\}$ to indicate that they are now used as measurements in the second MTT stage.

\subsubsection{MTT in 3-D}\label{sec:mtt3D} With TDOA estimates available across all sensors, odontocetess are tracked in 3-D. The 3-D tracking method performs multisensor data association and track initialization. At the time $k$, the state of the\vspace{.4mm} target $j$ is denoted as $\RV p_k^{(j)} = \big[\rv p_{k,x}^{(j)}\;\rv p_{k,y}^{(j)}\;\rv p_{k,z}^{(j)}\;\dot{\rv p}_{k,x}^{(j)}\;\dot{\rv p}_{k,y}^{(j)}\;\dot{\rv p}_{k,z}^{(j)}\big]^{\text T}$, where\vspace{-.2mm} $\rv p_{k,x}^{(j)}$, $\rv p_{k,y}^{(j)}$, and $\rv p_{k,z}^{(j)}$ are the position of the whale in a 3-D Cartesian coordinate system and $\dot{\rv p}_{k,x}^{(j)}$, $\dot{\rv p}_{k,y}^{(j)}$ and $\dot{\rv p}_{k,z}^{(j)}$ are the respective velocities.

The motion model is based on the\vspace{.5mm} kinematics
\begin{equation}
   \RV p_{k}^{(j)} = 
   \begin{bmatrix}
      \M I_3 & \hspace{-1.5mm} T_{M} \M I_3 \\[.7mm]
      \M 0_3 & \hspace{-1.5mm} \M I_3
   \end{bmatrix}
   \RV p_{k-1}^{(j)} +    \begin{bmatrix}
       \frac{T_{M}^2}{2} \M I_3 \\[.7mm]
       \M I_3
   \end{bmatrix} \RV w_{k}^{(j)}
\end{equation}
where $\RV w_{k}^{(j)}\in\mathbb{R}^3$ is a zero-mean Gaussian driving noise with covariance $\M I_3 \sigma_{\RV{w}}^2$ and driving noise std $\sigma_{\RV{w}}$.

Suppose that the target $j$ generated the TDOA $\hat{\rv{d}}^{(m)}_{k,s}$ at time $k$ and sensor $s$. The corresponding TDOA model is given \vspace{.5mm} by
\begin{equation}
   \hat{\rv{d}}^{(m)}_{k,s} = \big( \|\V p_k^{(j)} - \V{q}_{s_1}\| - \|\V p_k^{(j)} - \V{q}_{s_2}\| \big) / c + \rv{b}^{(m)}_{k,s} \label{eq:tdoa}
   \vspace{1mm}
\end{equation}
where $c$ is the speed of sound and $\rv{b}^{(m)}_{k,s}$ is zero-mean\vspace{0mm} Gaussian measurement noise with std $\sigma_{\rv{b}}$. The pdf that characterizes false positives,  $f_{\mathrm{fp}}^{(s)}\big(z_{k,s}^{(m)}\big)$, is again uniform on the interval $[-T_s^{\ist \text{max}} \hspace{1mm} T_s^{\ist \text{max}}]$.

Note that the nonlinear measurement model in Eq.~\ref{eq:tdoa} is underdetermined, i.e., the position of the whale is 3-D while the TDOA is only 1-D. Thus, for estimating the 3-D position, TDOAs provided by multiple sensors have to be fused. In particular, for fixed TDOA $\hat{d}^{(m)}_{k,s}$, Eq.~\ref{eq:tdoa} describes potential whale locations on a hyperboloid. If there were no MOU and no noise, one could estimate the 3-D whale location by computing the intersection of multiple hyperboloids provided by multiple TDOA sensors. However, due to the presence of MOU and noise and the number of whales being unknown, reliable state estimation can only be performed sequentially. In particular, we again use the MTT approach reviewed in Section \ref{sec:mttUnknown}. Due to the nonlinear measurement model Eq.~\ref{eq:tdoa}, we consider a particle-based implementation of this MTT approach. However, since the conventional particle filter suffers from weight degeneracy due to the curse of dimensionality\cite{DauHua:C03} and is thus only suitable for low dimensional states spaces, we here make use of the particle flow variant recently proposed in \onlinecite{ZhaMey:21}. 

Particle flow techniques are attractive methods that address the weight degeneracy issues\cite{DauHua:C07}. For each whale $j$ and time $k$, particles are migrated iteratively to the posterior distribution. Particle motion follows a stochastic process described by an ODE that expresses a Bayesian update step. The 
flow equations and the particles after the flow describe a proposal distribution that can be used for importance sampling \cite{LiCoa:J17}. By embedding particle flow into SPA-based MTT, the weight degeneracy is avoided, and MTT in 3-D from TDOA measurements can be performed with a reasonable number of particles and computational complexity \cite{ZhaMey:21}.

\section{Simulation}

An intrinsic challenge in tracking marine mammals using PAM is the lack of ground truth tracks; manual tracks can be erroneous since the data annotation process is subjective. Therefore, we set out to compare the 3-D tracking performance of different approaches from simulations to motivate the used 3-D tracking approach.

Four sets of 200 Monte Carlo simulations of whale tracks are generated and tracked using three different approaches: An approach (1) based on nonsequential tracking (NST) using a combination of DOAs described in \onlinecite{WigThaBauHil:R18}, (2) a single Bernoulli tracker\cite{RisVoVoFar:J13} (SBT), and (3) the proposed MTT approach described in Sec.\ref{sec:mtt3D}. In what follows, the approaches (1), (2), and (3) will be referred to as (NST), (SBT), and (MTT), respectively. Note that NST and SBT are genie-aided in the sense that they know the perfect association, the proposed MTT approach does not know the perfect association and performs data association automatically. Note that the perfect data association solution is only available if TDOA measurements are synthetically generated. In a real-world scenario, the perfect data association solution is unknown, and the genie has to be replaced by a human operator. Genie-aided SBT only uses the true measurements of a whale track as an input but uses the same clutter model and parameters as MTT.

Each set on Monte Carlo simulation has an increasing number of simultaneously present whales, ranging from 1 to 4. The whales' starting positions are 
placed uniformly on a circle of radius 1000 m and at a depth of 1000m. The birth distribution for MTT and SBT is assumed uniform on a region defined from -1000m to 1000m along the x- and y-axes and from -1800m to -500m along the z-axis.

 but a TDOA measurements are generated based on the same array geometry described in Sec. \ref{sec:data} and from the model described in Eq.~\ref{eq:tdoa} along with false alarms and missed detections. The same hyperparameters from Table \ref{tab:hyperparams} are used to generate the simulated data. Each simulation is 85 discrete time steps long, and the time step length is 7 s long. A whale is present for 50 time steps, and when there are multiple whales in the simulation, a whale is introduced every ten steps. In (NST) and (SBT), we use the TDOAs generated from the corresponding whales, i.e., we assume that there is a perfect human operator that can match the TDOA measurements to the correct whales.

\begin{figure}[t]
   \centering
   \includegraphics[width=\linewidth]{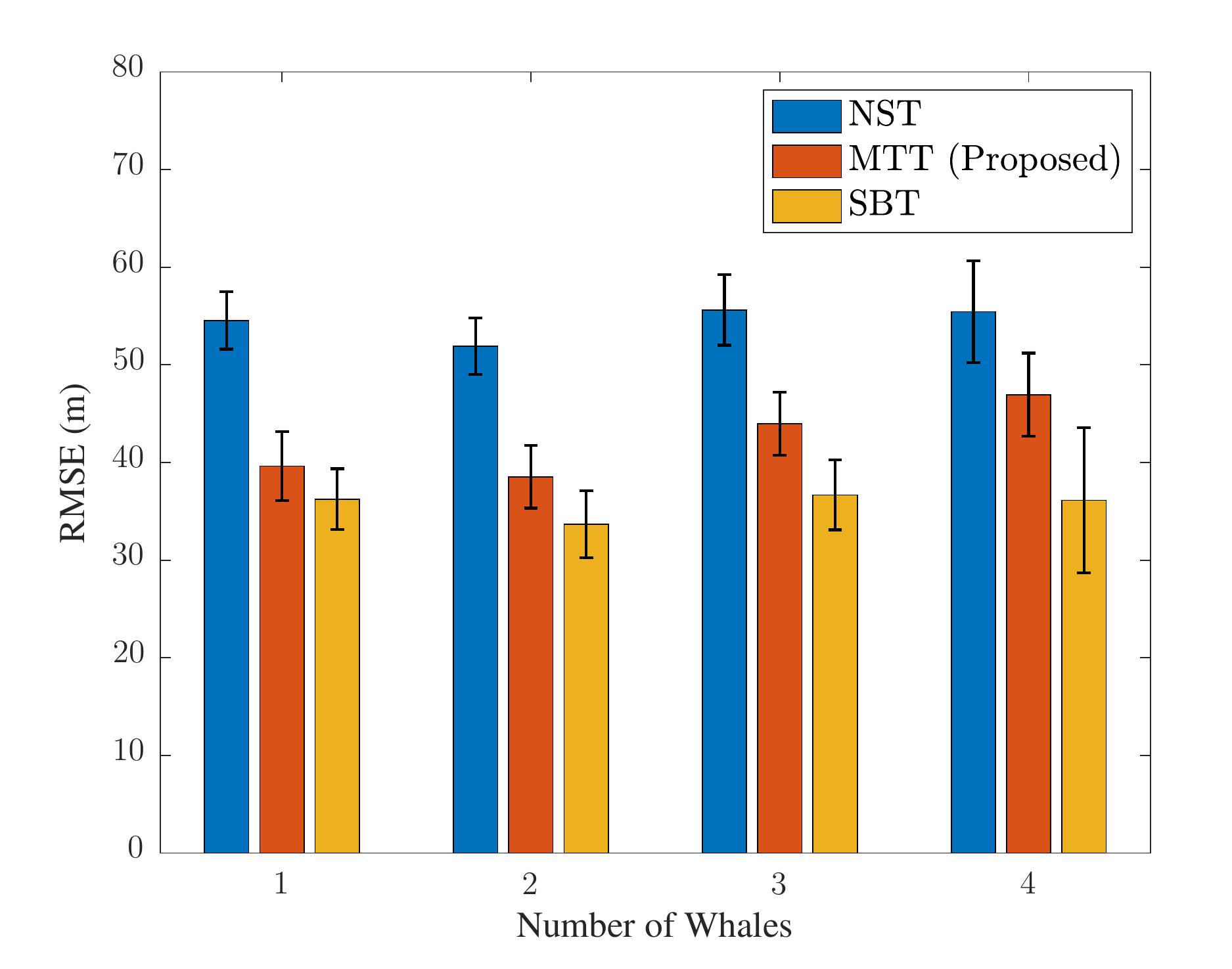}
   \caption{Bar graph of the simulated average RMSE versus simultaneously present whales for different 3-D whale tracking approaches. While (NST) and (SBT) are genie-aided in the sense that they know the perfect association, the proposed MTT approach does not know the perfect association and performs data association automatically. The error bars denote the 75th percentile of the measured RMSE across 200 Monte Carlo simulations.} 
   \label{fig:simRMSE}
\end{figure}

Each track's root-mean-square error (RMSE) is computed for each approach mentioned above (Fig.~\ref{fig:simRMSE}). The (MTT) and (SBT) sometimes do not detect the whale or have a significant error due to false alarms or missed detections. In that case, the RMSE is penalized by applying the error value of 110m, which is approximately twice the average RMSE of the (NST). From the simulation results, the (NST) has the highest RMSE since there is no filtering and the missed detections are interpolated. On the other hand, the (SBT) yields the lowest RMSE because the Bernoulli filter considers measurement noise and missed detection and filters the track accordingly. The (MTT) yields an RMSE between those from the (NST) and (SBT). We expect that the (MTT) will have an RMSE between those of (MTT) and (SBT). The (MTT) employs the Bernoulli filter but also solves the data association problem; hence, it is expected to perform worse than the (SBT). However, it performs filtering, so it is expected to perform better than the (NST). Furthermore, the RMSE increases with more whales because the data association problem becomes more challenging.

Sometimes a different number of whale tracks are generated by (MTT). Either a single track can be broken into two tracks, or an extra track is caused due to false alarms. The percentage of the number of simulations in which an extra number of tracks were generated was 0\% for one whale, 1\% for two whales, 11\% for three whales, and 19\% for four whale scenarios. Additional tracks can indeed be formed with the actual data, but in the final stage, either a human operator or an algorithm would join broken tracks or prune unlikely tracks.

\begin{figure}[t]
   \centering
   \includegraphics[width=\linewidth]{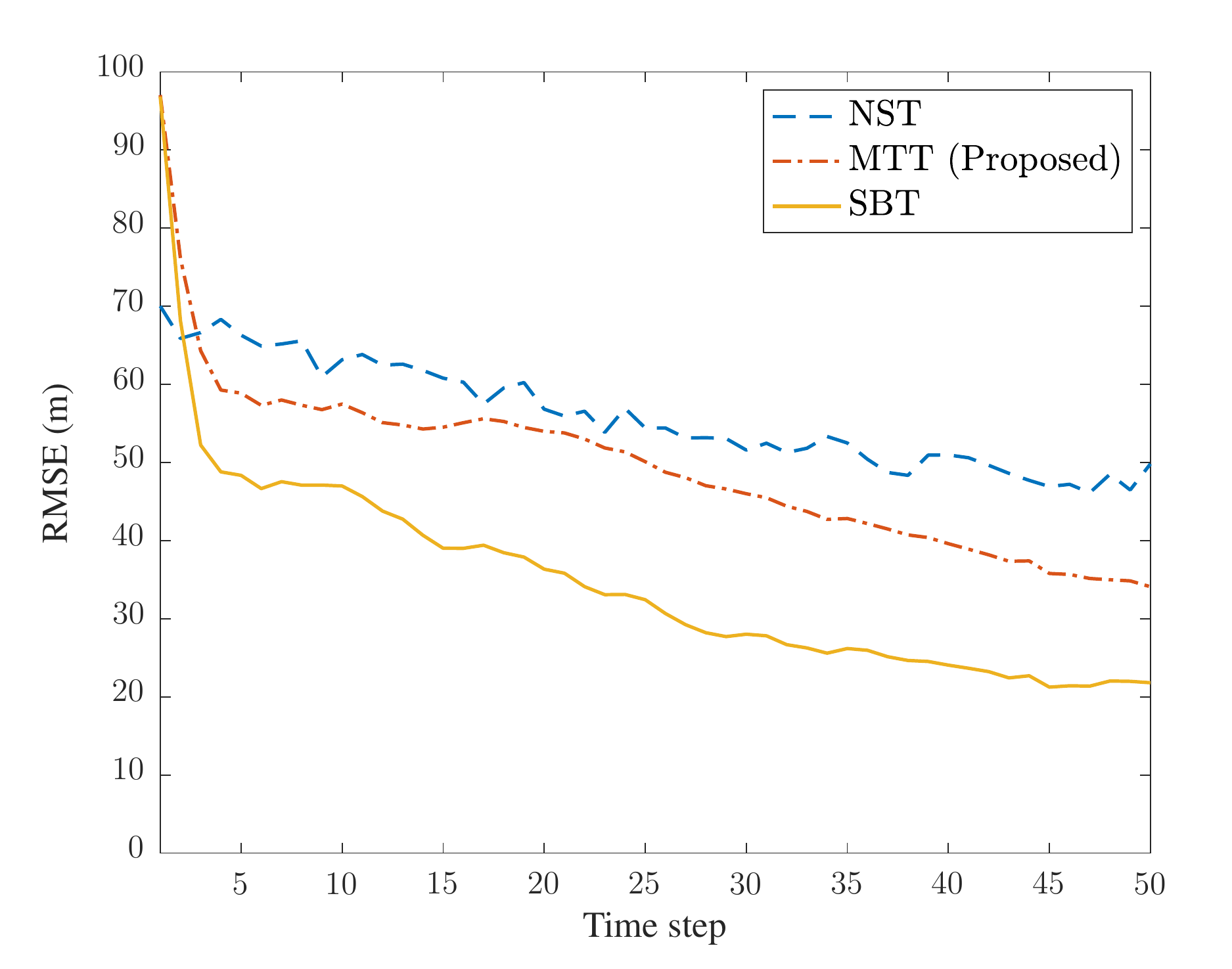}
   \caption{Average RMSE versus time for different 3-D tracking approaches. A scenario with four whales and 100 Monte Carlo simulation runs are considered. (SBT) and (MTT) perform sequential processing and thus converge to smaller error values over time.}
   \label{fig:simTrackRMSE}
\end{figure}

In Fig.~\ref{fig:simTrackRMSE}, the average RMSE over 800 tracks is shown as a function of the time step for the scenario with four whales. The RMSE of the initialization phase of (MTT) and (SBT) are high, but after a few steps, they converge to a smaller error value compared to the (NST). The actual data tracks are also longer than 50 time steps; therefore, we expect the total error to be even smaller for longer tracks, i.e., the real data. The simulation outcome shows that our approach can generate trustworthy whale tracks.

\section{Real Data Application}

The tracking capability of the proposed data processing chain is demonstrated on acoustic datasets containing echolocation clicks from Cuvier's beaked whales. In the signal processing step, their echolocation clicks are detected, and corresponding TDOA measurements among the pairs of hydrophones are computed using GCC-WIN described in Sec.~\ref{sec:gccImpl}. In the parameter estimation step, the whales are tracked first in the TDOA domain and then in the 3-D domain using the implementation of the MTT framework described in Sec.~\ref{sec:mttImpl}. The 3-D tracking results are compared to the tracks generated from hand-annotated DOA measurements following the approach in \onlinecite{WigThaBauHil:R18}.

\subsection{Data}\label{sec:data}

This study uses acoustic signals measured on two high-frequency acoustic recording packages (HARPs)\cite{WigHil:C07}, each of which is equipped with four hydrophones. The sampling frequency is 100 kHz; the corresponding Nyquist frequency is 50kHz, and the TDOA resolution is 10 \textmu s. The hydrophones on the HARP are 1m apart and arranged in a tetrahedral shape (see Fig.~2 in \onlinecite{GasWigHil:J15}) to form a small-aperture array. They were deployed off the coast of California (32\textdegree 39' 31.4''N, 119\textdegree 28' 37.6'' W) at a depth of $\sim$1330 m, and the arrays were approximately 1 km apart, establishing a large-aperture array. The arrays located east and west are referred to as HARP EE and HARP EW, respectively, and they recorded from March 15\textsuperscript{th}, 2018 to July 18\textsuperscript{th}, 2018, over 126 days. 

Since they share the same clock, the hydrophones on each HARP are time-synchronized. On the other hand, the large-aperture arrays are not synchronized. Nonetheless, as long as the echolocation clicks are measured on both HARPs within a time window in which the whale is considered stationary, precise synchronization is unnecessary\cite{Bag:J15,BarGriKliHar:J18}. 

In addition, the beaked whales' depth is far below the thermocline, with minimal change in the sound speed profile as a function of depth. Hence, we assume acoustic wave propagation with spherical spreading in an iso velocity medium, where the speed of sound is estimated as 1490m s\textsuperscript{-1}.

Five encounters of Cuvier's beaked whales have been processed, two of which are presented in this study, to demonstrate the tracking capability of the proposed data processing chain. In particular, we used data measured on June 11\textsuperscript{th} and July 1\textsuperscript{st} in 2018, which are 52 minutes and 20 minutes long, respectively. These encounters were detected using the long-term spectral average (LTSA) from the MATLAB-based program \textit{Triton}\cite{WigHil:C07}. 

Multiple species of whales could be present simultaneously near the PAM instruments, and their bioacoustic signals could interfere with one another. In such a case, the proposed data processing chain would be extended with classifying algorithms to discern among the species. However, there was no interference from other marine mammals in the datasets used, and thus the classification step was not required. Upon manual inspection, the echolocation clicks followed the characteristics of those from the Cuvier's beaked whales described in \onlinecite{BauMcDSimSolMerOleRocWigRanYacHil:J13}.

\subsection{Implementation}

\subsubsection{Signal Processing}\label{sec:sigProcImpl} 
\begin{figure}[t]
   \includegraphics[width=\linewidth]{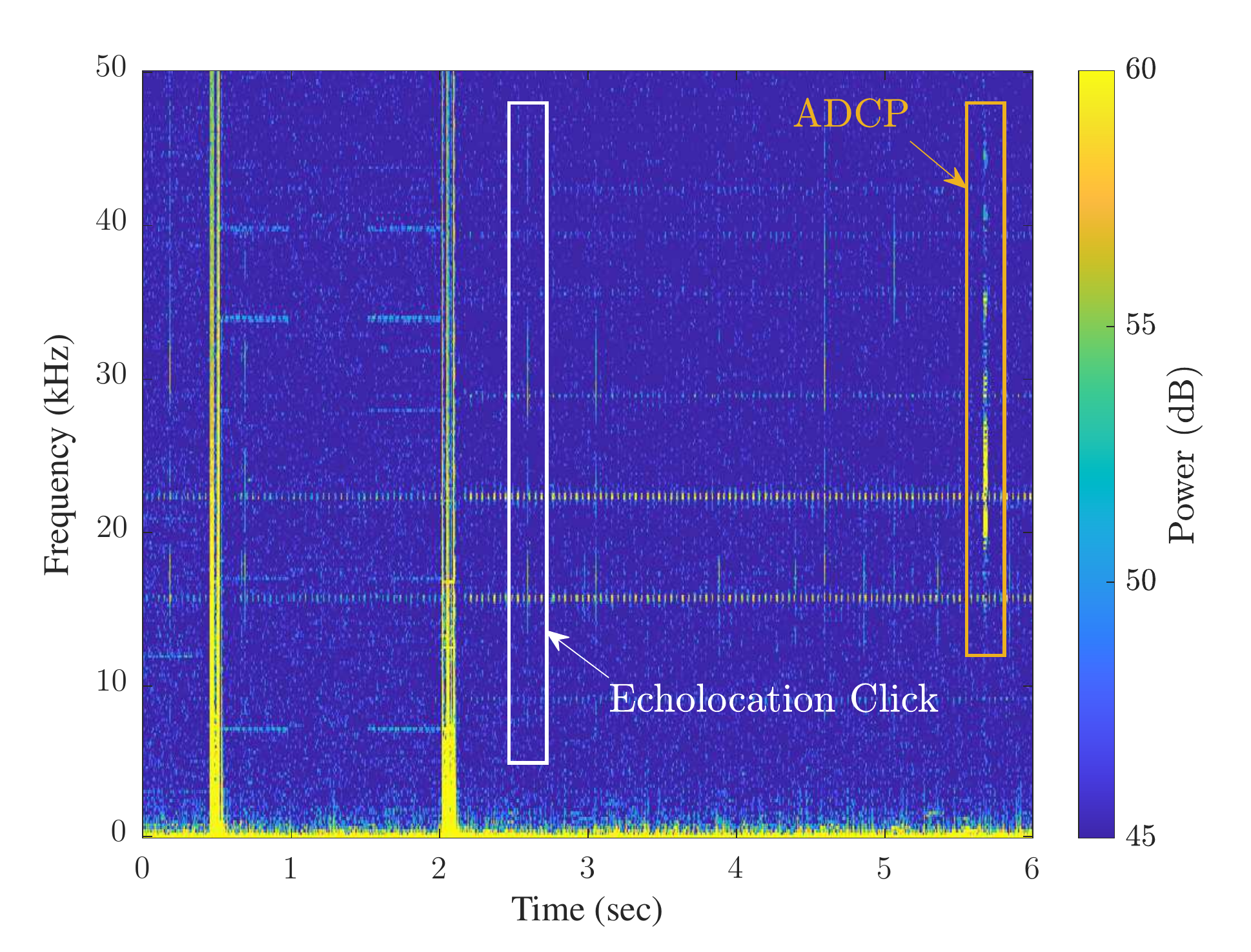}
   \caption{Example spectrogram of the data with echolocation clicks and the noise. Noise is produced by the ADCP, the instrument, and the undersea environment. The ADCP signal is the fast broadband signal with the center frequency near 25kHz labeled inside an orange box. The echolocation clicks are other broadband signals whose center frequencies are approximately 36 kHz. The instrument noise is a harmonic narrow band signal that is repeated every 31 s, whose spectrogram is displayed in Fig.~\ref{fig:noiseTemplate}.\vspace{-2mm}}
   \label{fig:signalExample}
\end{figure}

GCC-WIN is used to detect the echolocation clicks and estimate their TDOAs. Three noise sources are identified: a pulse signal from the co-present Acoustic Doppler Current Profiler (ADCP), an instrument noise on HARP highly correlated among hydrophone measurements, and ambient noise (Fig.~\ref{fig:signalExample}). Note also that the instrument noise is harmonic and broadband, thereby giving rise to multiple high amplitude peaks at wrong time delay locations when these signals are cross-correlated. This would hinder the tracker's performance significantly. The ambient noise was dominant below 13kHz, while the instrument noise was the major source of noise above 13kHz.

The signal was first pre-filtered with a high pass filter based on a Parks-McClellan optimal FIR filter design\cite{RabMccPar:J75}. The stopband frequency was at 13kHz, a passband frequency was at 15kHz, and the filter was applied with a zero-phase shifting digital filter\cite{OpeSha:B89}. Then, the ADCP signal was identified and removed. A nearby ADCP generated a short dominant pulse recurring approximately every 54 s, which completely flooded the acoustic measurements. Its center frequency was at 75kHz, but since the Nyquist frequency of the instrument was 50kHz, it was aliased and present at 25kHz. It was readily identified and nulled based on its high energy and center frequency characteristics. 

To perform GCC-WIN, a model of the noise is required. We observed that there was a pattern in the spectrogram of the instrument noise, which is the mechanical noise from the data storage system involving hard disks. This pattern repeats approximately every 31 s. An instrument noise spectrogram is estimated from a 20-minutes-long signal before the appearance of echolocation clicks in the data. This segment was chosen manually by inspecting the LTSA. The 20-minutes-long signal is further divided into shorter segments that are 31.65 s long. Their spectrograms are then averaged to estimate the noise model spectrogram (example in Fig.~\ref{fig:noiseTemplate}). All spectrograms are generated with FFT length of $N\rm=512$ samples, i.e. $T_{G} = 5.12$ ms (50\% overlap, Hamming window\cite{Har:J78}). Separate noise spectrograms were estimated for different days of data since the noise statistics varied throughout the deployment.

A critical step is to align the modeled noise spectrogram with the instrument noise in the signals of interest in the time domain. Upon applying a moving average filter to the signal of interest, indices at which the amplitudes are more significant than two times the std of the moving averaged signals are set to the mean amplitude from the signal of interest as they are likely to be the echolocation clicks and interfere with the alignment process. Then, the spectrogram of the signal of interest without any pulse-like alerts is computed using shorter segments of $N=512$ samples (50\% overlap, Hamming window\cite{Har:J78}). The GCC-SCOT amplitude at zero time lag is a measure of similarity between a specific snapshot of the demeaned noise spectrogram and that of the signal of interest. 

Once the noise model spectrogram and the signals are aligned, GCC-WIN is performed as described in Sec.~\ref{sec:gccImpl}. The resultant peaks from GCC-WIN, whose amplitudes are more significant than $P_{TDOA}=0.15$, are extracted and saved along with their amplitude information. When the GCC-WIN peak amplitude is more important than 10, i.e., detected a robust direct echolocation click, clicks within the next 40 ms are ignored since they are likely to be the reflections.

The TDOA measurements are merged and binned with a longer discrete time step length of $T_{M}=7$ and clustering distance $n_c=2$ samples follow the method described in Sec.~\ref{sec:gccImpl}. Given that the ICIs of the beaked whales range between 0.3 to 0.9 s\cite{GasWigHil:J15}, we can increase the probability of detection by inspecting over a longer time window. Moreover, seven seconds-long step length instead of 5.12 ms reduces the processing time by a factor of approximately 1,300. Since the average speed of the whale is 1.2 ms\textsuperscript{-1}, \cite{GasWigHil:J15}, it would have moved approximately 8.4 m in 7 s. Assuming that the whales are primarily hundreds of meters away from the arrays, the corresponding TDOA measurement is unlikely to change much over this period.

\subsubsection{Parameter Estimation} Since the TDOA is computed between a pair of hydrophones, there are $\binom{4}{2} = 6$ TDOA sensors per array and  $n_{\mathrm{s}}=12$ sensors total. As described earlier, the TDOA measurements are accumulated over $T_{M}\rmv=\rmv7$ s. The Cartesian coordinate system follows the East, North Up (ENU) convention: the x-, y-, and z-axes are positive along the East, North, and up directions. The depth is set to $z=0$ m at the sea surface, and the probability of the whale located at $z=0$ m is set to zero, i.e., $p_{k,z}^{(j)}=0$. The hyperparameters used for TDOA tracking and 3-D tracking are summarized in Table~\ref{tab:hyperparams}. For TDOA tracking, the birth distribution is chosen uniformly distributed between the minimum and maximum possible TDOA of a hydrophone pair. For 3-D tracking, the birth distribution is chosen uniformly distributed on the 3-D region of interest.

\begin{table}[h!]
\caption{\label{tab:hyperparams}Hyperparameters for tracking multiple beaked whales in TDOA and 3-D domains.} 

   	\begin{ruledtabular}
	\begin{tabular}{ccc}
      \textbf{Hyperparameters} & \textbf{TDOA Tracking} & \textbf{3-D Tracking}  \\ 
		\hline
	  Detection Probability, $p_d$                             & $0.80$                & $0.80$             \\
      Survival Probability, $p_{su}$                           & $0.90$                & $0.99$             \\ 
		Mean Number of False Positives, $\mu_{fp}$ & $10$       & $1$   \\
		Mean Number of Whale Birth, $\mu_{b}$ & $1.0\times 10^{-4}$  & $1$   \\
      Measurement Noise Std, $\sigma_{\RV u}$ \& $\sigma_{\RV w}$ & $1.0 \times 10^{-5}$  & $3.0 \times 10^{-5}$\\ 
      Driving Noise Std, $\sigma_{\RV v}$ \& $\sigma_{\RV b}$     & $1.5 \times 10^{-7}$  & $1.0\times 10^{-2}$\\
      Number of Particles                                      & $30,000$              & $100,000$          \\
      Minimum track length                                     & $20$                  & $5$                \\
   	\end{tabular}
	\end{ruledtabular}
\end{table}

A final pruning step is used to remove extra and unreasonable tracks. The study in \onlinecite{BarGriKliHar:J18} estimates that the average swim speed of a Cuvier's beaked whale is 1.2m s\textsuperscript{-1}, and \cite{GasWigHil:J15} showed that the horizontal speed could range from 1 to 3m s\textsuperscript{-1}. Consequently, if the median speed (Euclidean norm of the estimated velocity) of the whale track is more significant than 2m s\textsuperscript{-1} or the track length is shorter than five time steps, the track is discarded. Furthermore, at every time step, if its estimated speed is faster than 3.5m s\textsuperscript{-1}, the state at that time step is ignored.

\subsection{Results}
The results are compared to the tracking results from hand-annotated data using the framework proposed in \onlinecite{WigThaBauHil:R18}. In \onlinecite{WigThaBauHil:R18}, the detected TDOAs are used to compute the azimuth and elevation angles of the beaked whale relative to each hydrophone. The whales are tracked manually in the azimuth and elevation angles domain, whose tracks are fused between two hydrophones to estimate the 3-D locations of the whale.
 
\begin{figure}[t]
   \centering
   \includegraphics[width=0.75\linewidth]{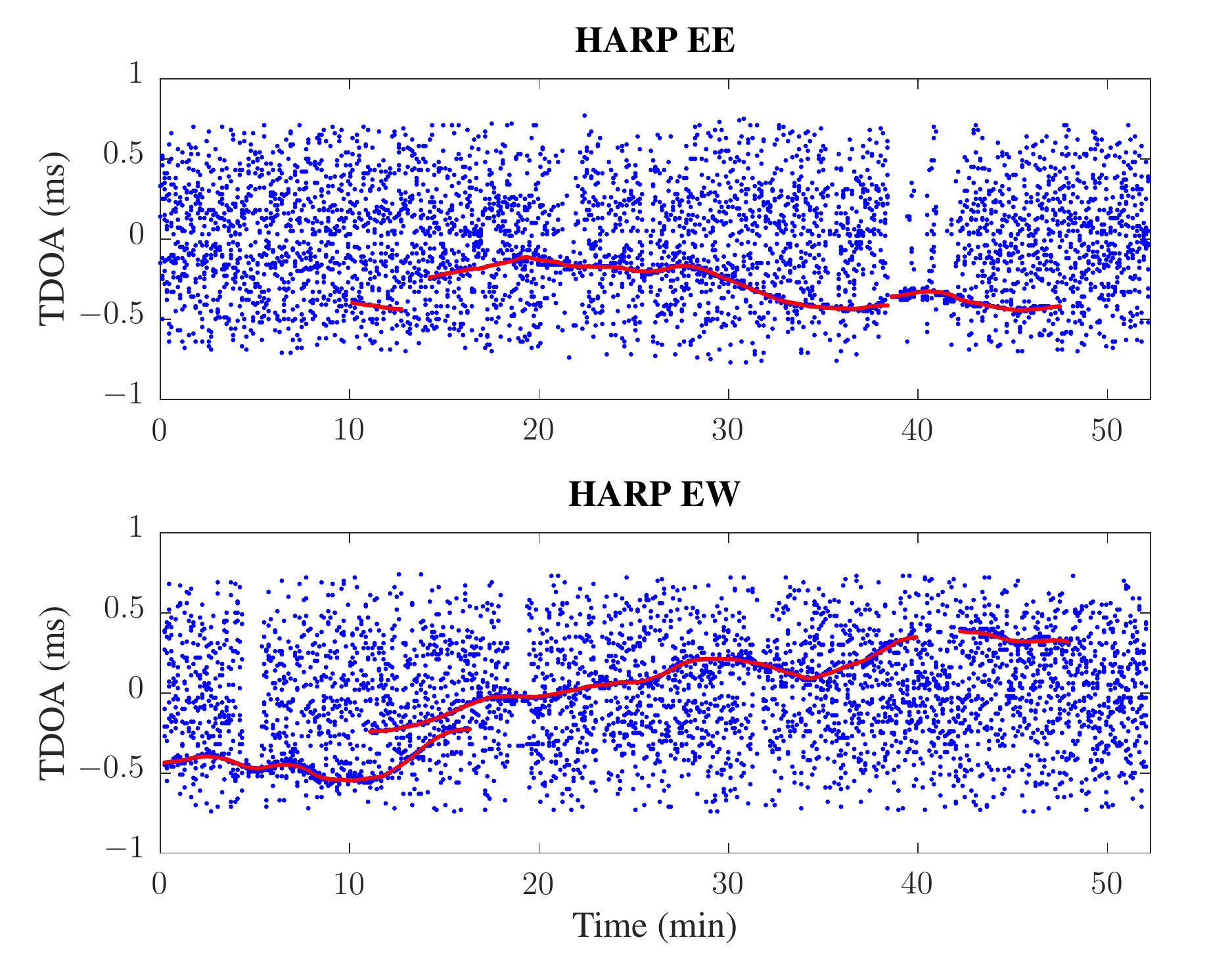}
   \caption{Example TDOA measurements (blue dots) and corresponding TDOA tracking results (red lines) from one sensor of HARP EE (top) and one sensor of HARP EW (bottom). Two whales were observed by HARP EW, while only one was observed by HARP EE. The 3-D tracking result in Fig.~\ref{fig:results0611} shows that both whales were much closer to HARP EW. The considered acoustic data were collected on June 11\textsuperscript{th}, 2018.}
   \label{fig:tdoa0611}
\end{figure}

\begin{figure}[t]   
\centering
\includegraphics[width=\linewidth]{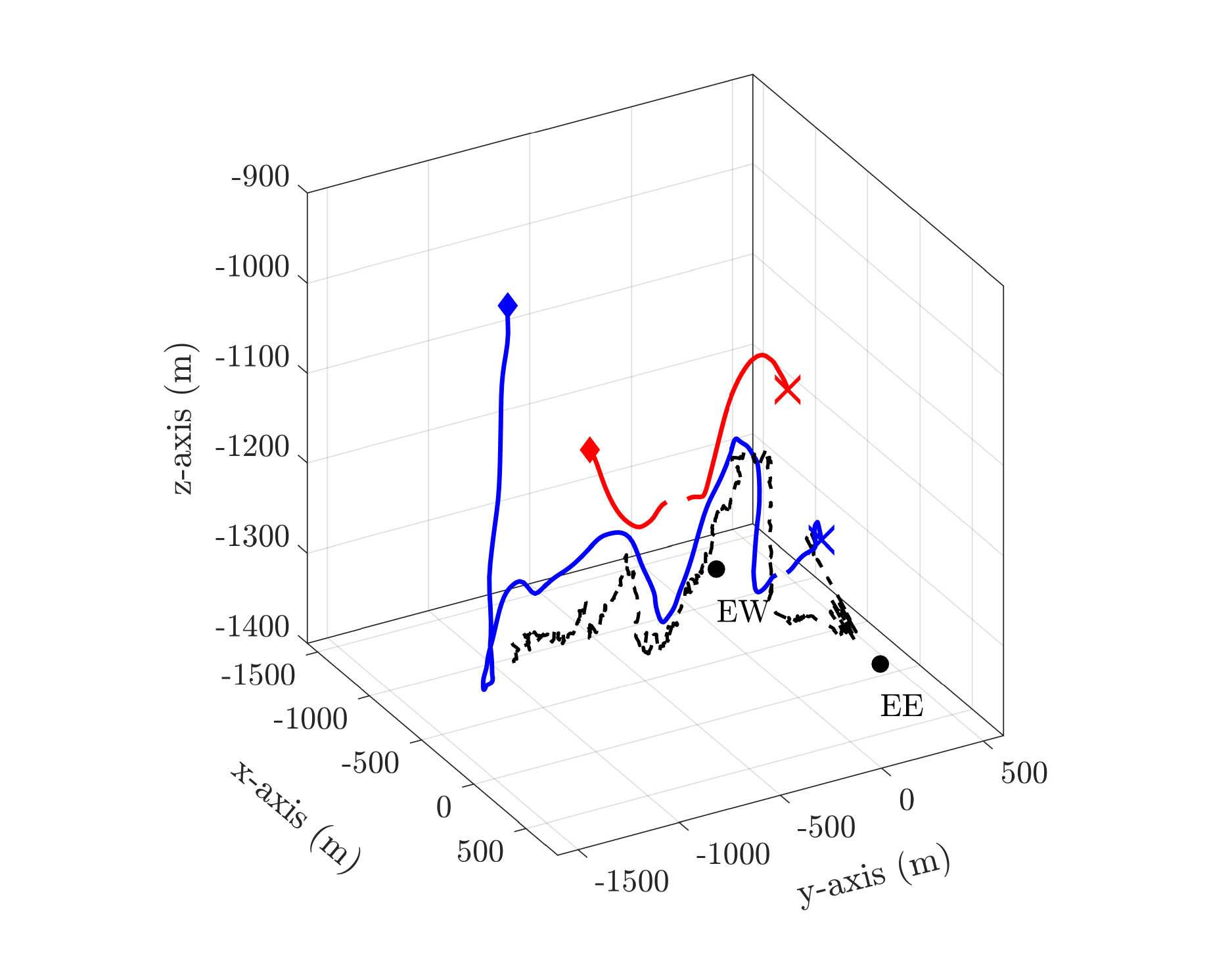}
\caption{Comparison of the hand-annotated track (dashed black line) and the multi-target tracking results (blue and red lines) using the data from June 11\textsuperscript{th}, 2018. The diamond and the cross indicate each track's start and end, respectively. The blue track is generated from TDOA measurements, which mostly originated from the EW array (Fig.~\ref{fig:tdoa0611}). There is a short gap in the red track since parts of the track that corresponded to a biologically unfeasible speed were removed. Note that a new track was found and that the diving behavior of a whale, which a human operator did not capture, was extracted.}
\label{fig:results0611}
\end{figure}

\begin{figure}[t]   
   \centering
   \includegraphics[width=\linewidth]{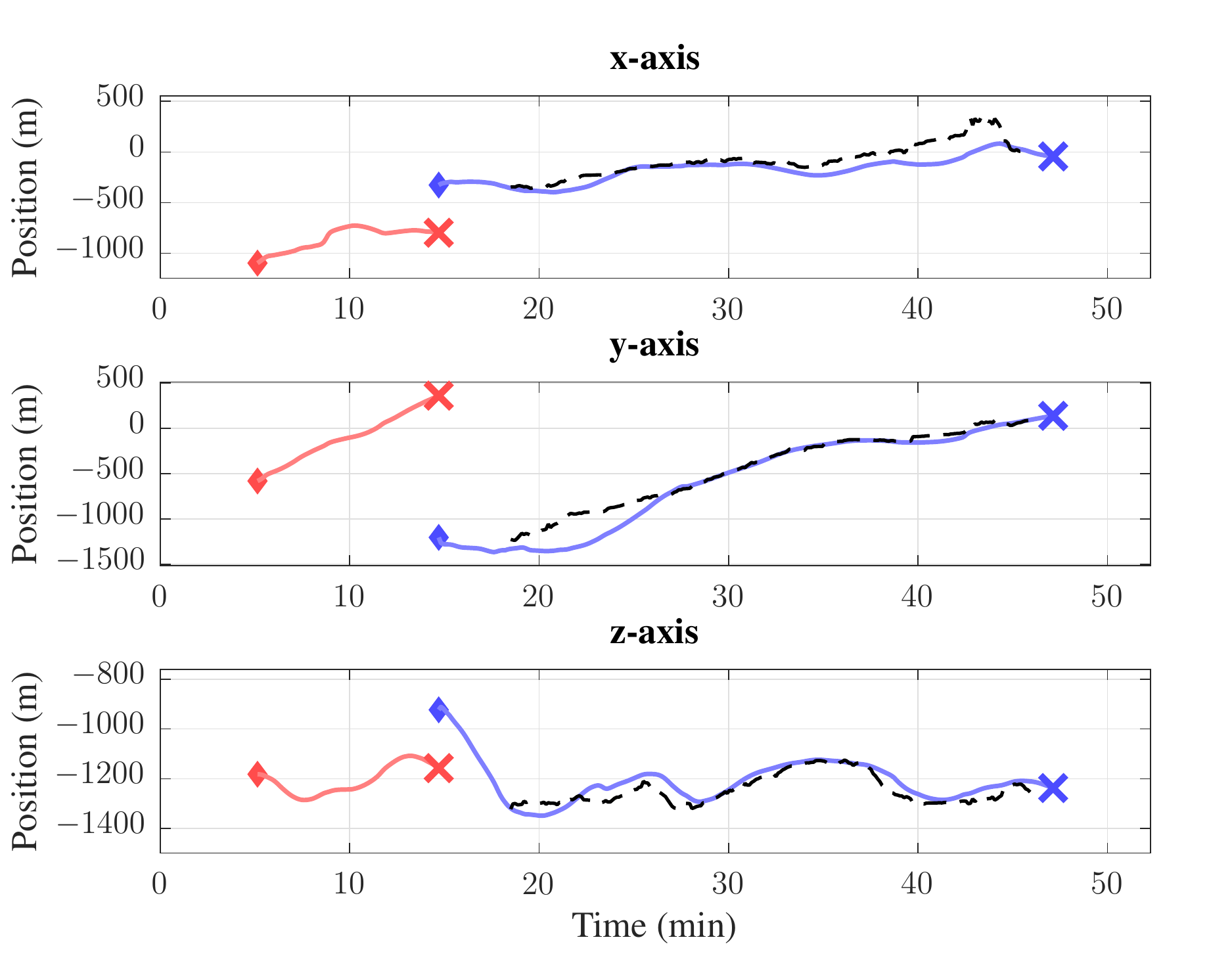}
   \caption{Tracks of the beaked whales from the data recorded on June 11\textsuperscript{th}, 2018. The hand-annotated track are shown as dashed black lines, and the multi-target tracking results are shown as blue and red lines.}
   \label{fig:results0611Exploded}
\end{figure}

We present the tracks in the TDOA and 3-D domains of two different datasets. Our tracking algorithm identified and tracked two whales in the data recorded on June 11\textsuperscript{th} 2018. Even though HARP EE recorded echolocation clicks from only one whale, HARP EW detected those from two whales (Fig.~\ref{fig:tdoa0611}). Based on the 3-D tracks (Fig.~\ref{fig:results0611} and Fig.~\ref{fig:results0611Exploded}), the first whale (blue line) was far from HARP EE, hence the lack of detected echolocation clicks on HARP EE. To verify further, the corresponding TDOA measurements from the first whale were computed using the TDOA model, which was in accordance with the tracked TDOA. The manual tracking method could not generate the track for the first whale because it requires that the DOAs exist on both arrays simultaneously.

\begin{figure}[t]
   \centering
   \includegraphics[width=0.75\linewidth]{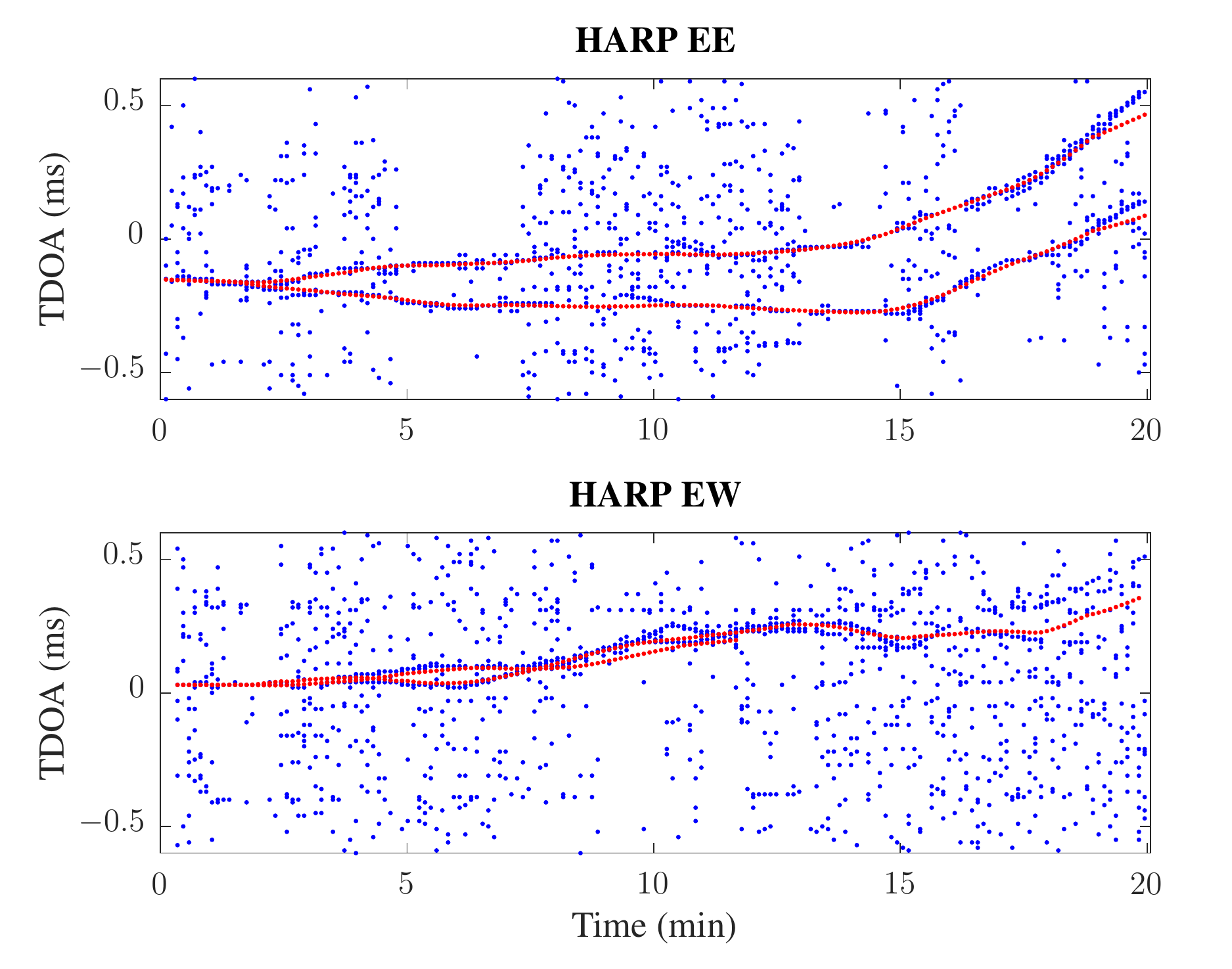}
   \caption{Example TDOA measurements (blue dots) and corresponding TDOA tracking results (red lines) from one sensor of HARP EE (top) and one sensor of HARP EW (bottom). One of the whales was not successfully tracked towards the end by HARP EW  because of long gaps of missing TDOA. The considered acoustic data were collected on July 1\textsuperscript{st}, 2018. \vspace{5mm}}
   \label{fig:tdoa0701}
\end{figure}

\begin{figure}[t]
   \includegraphics[width=\linewidth]{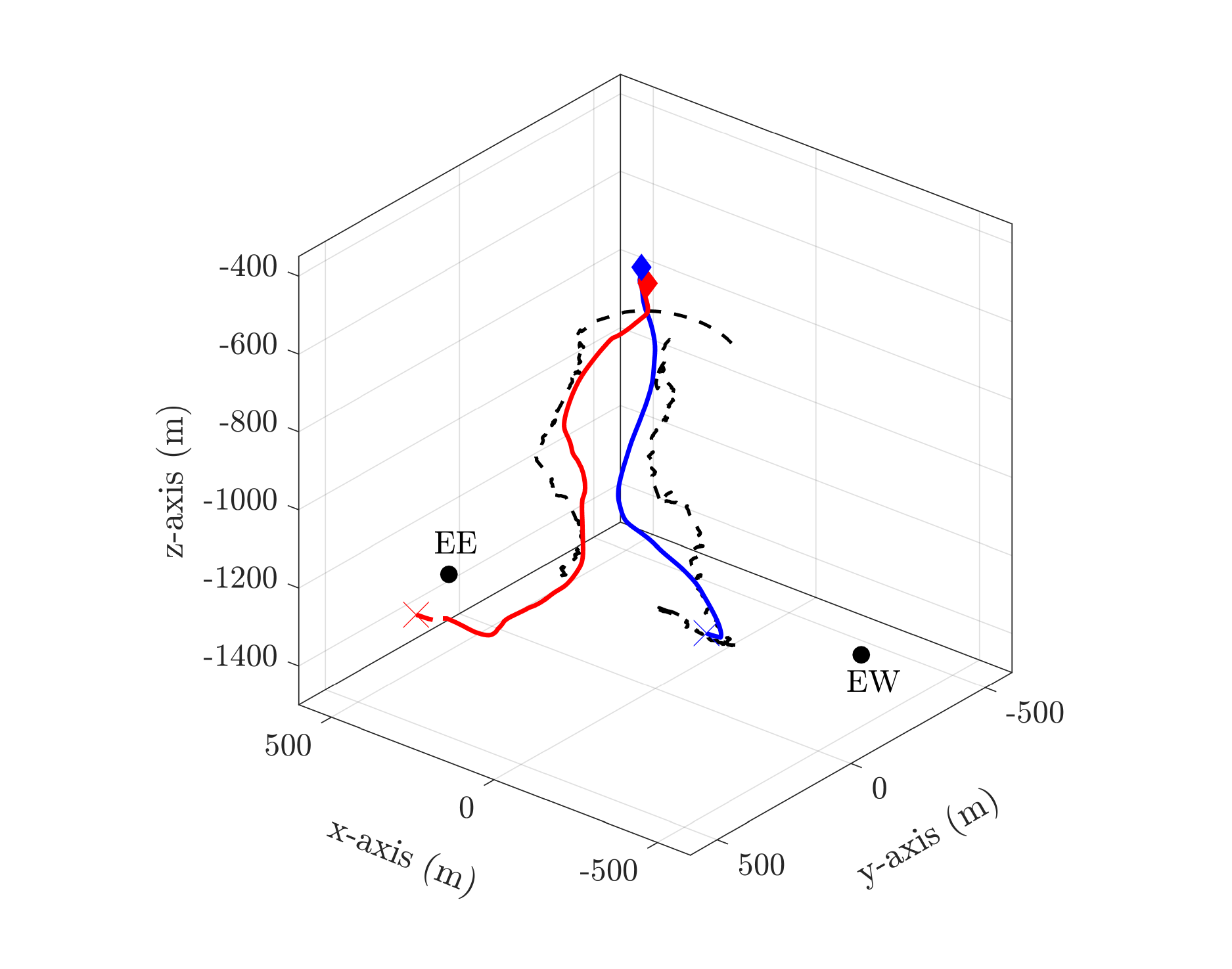}
   \caption{Comparison of the hand-annotated track (dashed black lines) and the multi-target tracking results (blue and red lines) using the data from July 1\textsuperscript{st}, 2018. The diamond and the cross indicate the start and the end of each track, respectively. Two whales are simultaneously diving into deeper waters.}
   \label{fig:results0701}
\end{figure}

\begin{figure}[t]
   \centering
   \includegraphics[width=0.75\linewidth]{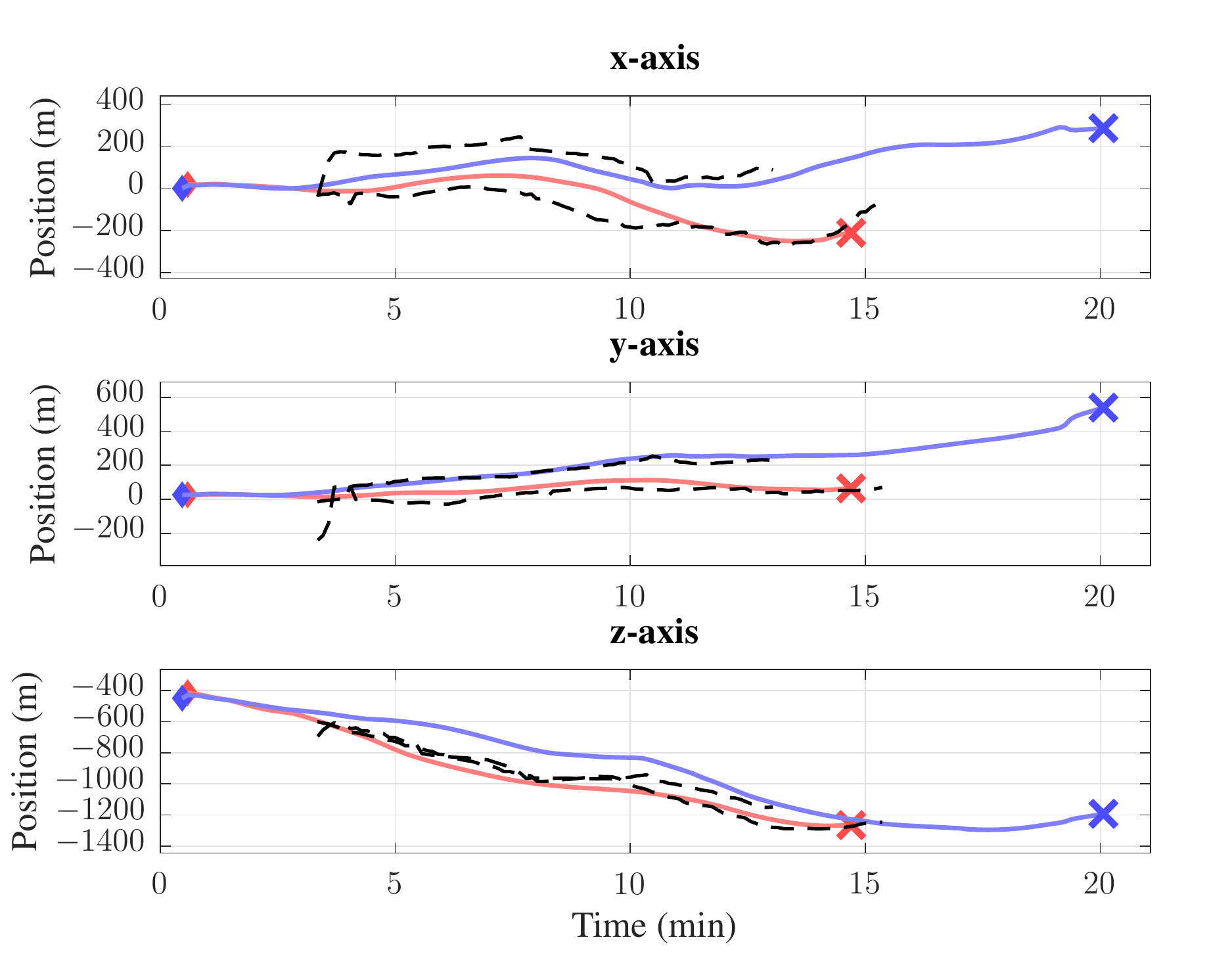}
   \caption{Tracks of the beaked whales from the data recorded on July 1\textsuperscript{st}, 2018. The hand-annotated track are shown as dashed black lines, and the multi-target tracking results are shown as blue and red lines.}
   \label{fig:results0701Exploded}
\end{figure}

Again, two whales were present in the data collected on July 1\textsuperscript{st}, 2018 (Fig.~\ref{fig:tdoa0701}, Fig.~\ref{fig:results0701}, and Fig.~\ref{fig:results0701Exploded}). Two whales were diving together from shallow depths, where the initial locations are at a depth of approximately 500m. Two tracks were close to each other, but the tracker was capable of separating the two tracks. By using GCC-WIN, more echolocation clicks were detected such that longer 3-D tracks (approximately 5 minutes long) were generated with our approach. In addition, while the tracks are similar between the manually tracked results and our results, the blue track and its corresponding manual track are noticeably different at the shallower depth. To verify the reliability of our results, the tracker was run with the data in the reverse time steps. Since it is easier for the tracking algorithm to identify two whales if they are spatially distinct, the tracks from the reverse order would be more reliable in this scenario. We confirmed that the correct and reverse order tracking results were identical. 

\section{Discussion}
A data processing chain for automatically detecting and tracking multiple odontocetes from their echolocation clicks was developed. It has successfully tracked multiple Cuvier's beaked whales in 3-D from their echolocation click recordings made on a pair of volumetric hydrophone arrays. No human operators or heuristics were needed to initialize the tracks and combine the TDOA measurements corresponding to individual whales. The graph-based MTT method efficiently solves the data association problem to fuse the TDOA measurements among missed detections and false positives and track odontocetes in a computationally tractable manner. In addition, we demonstrated that new and more extended tracks of the beaked whales could be extracted using GCC-WIN, which normalizes the cross PSD by the estimated instrument noise PSD to whiten the instrument noise. 

A few things need to be considered when adapting the processing chain for applications with another set of bioacoustic data from whales. As long as it is given acoustic data of echolocation clicks of any odontocetes, the data processing chain is species-agnostic since the tracks are distinguished based on spatial information. The pre-processing step needs careful customization, e.g., $T_G$, frequency filters, etc., so echolocation clicks from species of interest are processed. A manual or automated classification step would be desirable for scientific purposes. In addition, GCC-WIN is only recommended if the instrument noise is dominant and can be estimated; otherwise, other GCC adaptations, such as SCOT or PHAT, are more efficient methods for TDOA estimations. Moreover, if the noise is uncorrelated or weakly correlated, computing TDOAs from click trains would be more suitable than accumulating and clustering individual echolocation clicks.  

As seen in the second scenario (Fig.~\ref{fig:results0701}), closely spaced tracks, i.e., track coalescence, can pose a challenge. To mitigate, the future model in the MTT framework could incorporate statistics of ICIs to distinguish among multiple whales. It is observed that the clustering described in Sec.~\ref{sec:gccImpl} to ensure the assumption of a single measurement per target has hindered the track results under coalescence. 

In addition, it is known that some species of odontocetes are highly sociable and thus sometimes move together in close spatial proximity\cite{ConManTyaWhi:J98,McHugh:B19}. In such a case, it would be challenging to track an individual whale, but subgroups could be tracked\cite{GruNosOle:J21}. The question of the feasibility of the MTT frameworks to follow pods of odontocetes instead of individuals poses an interesting research problem. 

Another challenge is that the model based on a constant probability of detection is not entirely reflective of reality. Even though the likelihood of detection is affected by the SNR of the bioacoustic signal, their irregularity, i.e., occurrences of a burst of clicks followed by long gaps of silence, needs to be considered. For example, there is approximately a two-minutes-long gap of TDOA detections starting at 40 minutes in HARP EW in Fig.~\ref{fig:tdoa0611} likely due to the beaked whale looking away from the array. With prior information on the echolocation click directionality and the motion of the odontocetes at each given time, the detector's performance could be modelled more accurately. Alternatively, a time-varying detection probability could be estimated together with the target states\cite{SolMeyBraHla:J19}.

\section{Conclusion}
In this paper, we developed a data processing chain for the fully automated detection and tracking of odontocetes in 3-D from echolocation clicks. The detection rate of the echolocation clicks is improved by utilizing a generalized cross-correlation algorithm designed to whiten instrument noise. Multiple whales are detected and tracked simultaneously by applying two stages of a graph-based tracking method that efficiently solves the data association problem. Graph-based detection and tracking are first performed for each hydrophone pair individually in the TDOA domain and subsequently in the 3-D domain.  The ability to track multiple odontocetes simultaneously without manual data selection by a human operator has been demonstrated based on real acoustic data provided by two volumetric hydrophone arrays. In particular, we presented tracking results in a scenario with two echolocating Cuvier's beaked whales (\textit{Ziphius cavirostris}). These results show how the proposed data processing chain can simplify scientists' steps to study the deep-diving echolocating odontocetes. In addition, our simulation results suggest that the presented processing chain can be used to track a larger number of whales in a scalable manner and is thus particularly appealing for future passive acoustic monitoring (PAM) systems. 

\begin{acknowledgments}
This research was supported in parts by the National Science Foundation (NSF) under CAREER Award N2146261, by the Office of Naval Research (ONR) under YIP Award N00014-15-1-2587, and by the Cooperative Ecosystems Study Unit under Cooperative Agreement N62473-18-2-0016 for the U.S. Navy, U.S. Pacific Fleet. We thank the students and staff of the Scripps Marine Bioacoustics Collaborative for fieldwork, data curation, and data preparation.
\end{acknowledgments}

\bibliography{Manuscript}

\end{document}